\documentclass[12pt]{article}
 \usepackage{graphicx}
\RequirePackage{amsthm, amsmath, amsfonts, amssymb}
\RequirePackage{mathtools}
\RequirePackage[authoryear]{natbib}
\defcitealias{song2020score}{Song et al. 2020a}
\defcitealias{song2020improved}{Song et al. 2020b}
\defcitealias{song2020denoising}{Song et al. 2020c}
\defcitealias{chen2023score}{Chen et al. (2023a)}
\defcitealias{chen2023efficient}{Chen et al. 2023b}
\RequirePackage[pdfencoding=auto,psdextra,colorlinks,citecolor=black,urlcolor=blue]{hyperref}
\RequirePackage{epsfig,longtable,xcolor,algpseudocode}
\RequirePackage{graphicx}

\usepackage{xr}
\usepackage[linesnumbered, ruled, vlined]{algorithm2e}
\usepackage{multirow}
\usepackage{makecell}
\usepackage{subcaption}
\usepackage{booktabs}

\newcommand{\blind}{1}

\allowdisplaybreaks

\newcommand{\bfm}[1]{\ensuremath{\mathbf{#1}}}
\def\ba{\bfm a}     \def\bA{\bfm A}     \def\cA{{\cal  A}}

\def\be{\bfm e}          \def\cE{{\cal  E}}     
\def\bff{\bfm f}         \def\cF{{\cal  F}}     
     \def\bG{\bfm G}          
          \def\cH{{\cal  H}}     
     \def\bI{\bfm I}

          \def\cL{{\cal  L}}     
          \def\cM{{\cal  M}}     
          \def\cN{{\cal  N}}     
               
          \def\cP{{\cal  P}}

\def\bs{\bfm s}

          \def\cV{{\cal  V}}     
     \def\bW{\bfm W}          
\def\bx{\bfm x}     \def\bX{\bfm X}     \def\cX{{\cal  X}}     
               
\def\bz{\bfm z}     \def\bZ{\bfm Z}     \def\cZ{{\cal  Z}}

\newcommand{\bfsym}[1]{\ensuremath{\boldsymbol{#1}}}
\def \balpha   {\bfsym{\alpha}}

         \def \btheta   {\bfsym{\theta}}
        
      \def \bmu      {\bfsym{\mu}}

\def \bGamma   {\bfsym{\Gamma}}

\DeclareMathOperator*{\argmin}{argmin}

\DeclareMathOperator{\diag}{diag}

\DeclareMathOperator{\supp}{supp}

\def \RR	{\mathbb{R}}

\def \HH	{\mathbb{H}}
\def \PP {\mathbb{P}}

\def \EE {\mathbb{E}}
\def \ZZ {\mathbb{Z}}

\def \cV {\mathcal{V}}

\def \tilde{\widetilde}

\newcommand{\beq}  {\begin{equation}}
\newcommand{\eeq}  {\end{equation}}
\newcommand{\beqn} {\begin{eqnarray}}
\newcommand{\eeqn} {\end{eqnarray}}
\newcommand{\beqnn}{\begin{eqnarray*}}
\newcommand{\eeqnn}{\end{eqnarray*}}

\numberwithin{equation}{section}
\theoremstyle{plain}
\newtheorem{thm}{Theorem}[section]
\newtheorem{lem}{Lemma}[section]

\newtheorem{defn}[thm]{Definition}
\newtheorem{prop}{Proposition}[section]
\newtheorem{cond}{Condition}[section]

\newtheorem{exmp}[thm]{Example}
\newtheorem{rem}{Remark}[section]

\begin{document}

\def\spacingset#1{\renewcommand{\baselinestretch}%
{#1}\small\normalsize} \spacingset{1}


\if1\blind
{
  \title{\bf Denoising Diffused Embeddings: a Generative Approach for Hypergraphs}
 \author{Shihao Wu, Junyi Yang, Gongjun Xu, and Ji Zhu \\
    Department of Statistics, University of Michigan
    }
    \date{}
  \maketitle
} \fi

\if0\blind
{
  \bigskip
  \bigskip
  \bigskip
  \begin{center}
    \title{\bf Denoising Diffused Embeddings:\\ Efficient and Interpretable Generative Modeling for Hypergraphs}
\end{center}
  \medskip
} \fi

\bigskip
\begin{abstract}
Hypergraph data, which capture multi-way interactions among entities, are increasingly prevalent in the big data era. Generating new hyperlinks from an observed, usually high-dimensional hypergraph is an important yet challenging task with diverse applications in areas such as electronic health record analysis and biological research. This task is fraught with several challenges. The discrete nature of hyperlinks renders many existing generative models inapplicable. Additionally, powerful machine learning-based generative models often operate as black boxes, providing limited interpretability.
Key structural characteristics of hypergraphs, including node degree heterogeneity and hyperlink sparsity, further complicate the modeling process and must be carefully addressed.
To tackle these challenges, we propose Denoising Diffused Embeddings (DDE), a general and efficient generative modeling  architecture for hypergraphs. DDE exploits low-rank structure in high-dimensional hypergraphs via a conditional hyperlink likelihood model that links discrete hyperlinks to a continuous latent embedding space and leverages a score-based diffusion model to reconstruct that space.
Theoretically, we show that when true latent embeddings are accessible, DDE exactly reduces the task of generating new high-dimensional hyperlinks to generating new low-dimensional embeddings. Moreover, we analyze the implications of using estimated embeddings in DDE, revealing how hypergraph characteristics such as dimensionality, node degree heterogeneity, and hyperlink sparsity impact its generative performance.
Simulation studies demonstrate the superiority of DDE over existing methods, in terms of both computational efficiency and generative performance. Furthermore, an application to a symptom co-occurrence hypergraph derived from electronic medical records uncovers interesting findings and highlights the advantages of DDE. 
\end{abstract}

\noindent%
{\it Keywords:}  hypergraphs, generative models, latent embeddings, diffusion models.
\vfill

\newpage
\spacingset{1.8} 

\section{Introduction}

Hypergraph data, which capture multi-way interactions among entities, are increasingly prevalent in the big data era. For example, the Medical Information Mart for Intensive Care dataset \citep[MIMIC-III;][]{johnson2016mimic} includes clinical records from 49,785 hospital admissions involving 38,597 unique adult patients. These records document the co-occurrence of clinical symptoms and procedures within patient profiles, capturing the underlying relationships among medical symptoms and procedures.  Similarly, in protein interaction datasets \citep{rhodes2005probabilistic, nepusz2012detecting}, groups of proteins function together in metabolic reactions, while in gene interaction datasets \citep{rinaldo2005characterization,shojaie2009analysis}, groups of genes associate with each other to potentially form protein complexes. These interactions, which can be represented by hypergraphs,
provide valuable insights into the functional roles of genes and proteins in biological processes.

\begin{figure}[h]
      \centering
      \vspace*{0.1cm}
      \includegraphics[height=3.6cm]{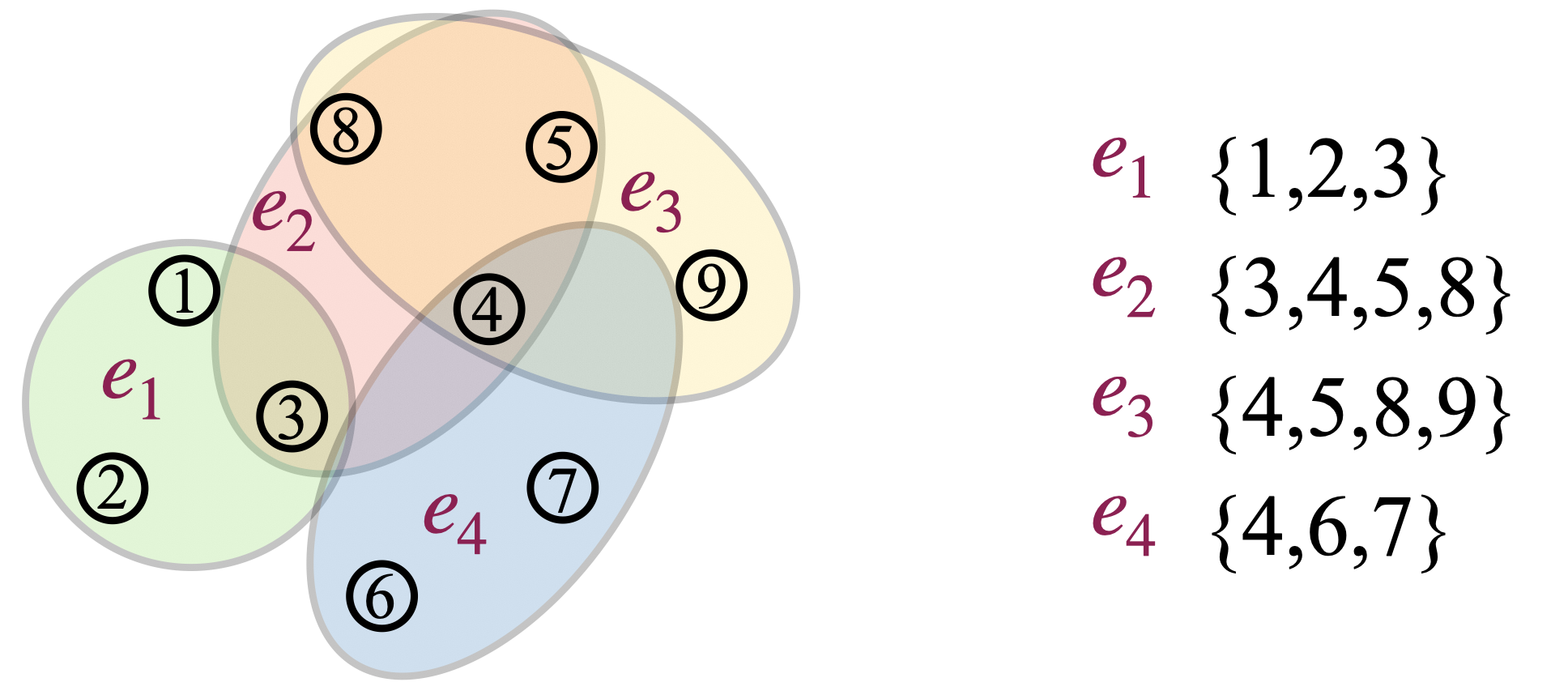} 
      \caption{Example of an observed hypergraph $\cH([9],\{e_1,e_2,e_3,e_4\})$. The hyperlinks observed from the hypergraph in the left figure are shown as a multiset of sets on the right.} 
            \vspace*{-0.3cm}
      \label{fig:hypergraph}
\end{figure}

A hypergraph can be denoted by $\cH(\cV_n,\cE_m)$, which consists of the node set $\cV_n = \{v_1,\ldots,v_n\}$ and a collection of hyperlinks/hyperedges $\cE_m = \{e_1,\ldots,e_m\}$. For simplicity, we denote $\cV_n = [n] =\{1,\ldots,n\}$. Each hyperlink is a subset of $[n]$, representing the nodes on that hyperlink. For example, if $e_1 = \{1,2,3\}$, then nodes 1,2 and 3 form hyperlink 1. Figure \ref{fig:hypergraph} illustrates an example of an observed hypergraph with $9$ nodes and $4$ \emph{observed} hyperlinks. 
Hyperlinks have an \emph{order}, defined as the number of nodes they contain. In Figure \ref{fig:hypergraph}, the four hyperlinks have orders of 3, 4, 4, and 3, respectively. Moreover, the nodes in a hypergraph typically exhibit \emph{heterogeneous degrees}, reflecting their varying ``popularity''. For example, in Figure \ref{fig:hypergraph}, node $4$ appears $3$ times among the four observed hyperlinks, indicating a higher degree than node $2$, which appears only once. Besides node degree heterogeneity, another characteristic of real-world hypergraphs is the sparsity of hyperlinks: the orders of the hyperlinks are usually much smaller in comparison to the total number of nodes. For instance, in the MIMIC-III dataset, a symptom co-occurrence hypergraph can be constructed where medical symptoms are nodes, and each patient profile forms a hyperlink, with nodes representing the symptoms exhibited by the patient. This hypergraph contains 6,985 nodes and 49,785 hyperlinks, with median and mean hyperlink orders of 9 and 12.6, respectively, highlighting the sparsity of the hyperlinks relative to the total number of nodes. Moreover, the node degrees in this hypergraph are highly heterogeneous, as indicated by the 25th and 75th percentiles of node frequencies, which are 2 and 27, respectively.

In this work, we study generative models for hypergraphs, and aim to generate new hyperlinks based on an observed hypergraph. 
Synthetic hyperlink generation has significant applications in modern data analysis. 
For instance, in electronic health record (EHR) data, symptom co-occurrences in patient profiles can be represented as hyperlinks. 
Since each medical center typically has a limited number of patient profiles, deeper and more accurate analyses often require data sharing across centers. However, 
directly sharing these hyperlinks raises privacy concerns. Generating synthetic hyperlinks that closely match the distribution to real ones while remaining distinct offers a practical solution. These synthetic hyperlinks enable data pooling across medical centers, facilitating richer analyses while avoiding the direct sharing of patient information.       
Other examples of applications include generating new combinations of protein or gene subgroups that may function together, creating product bundles for online shoppers, and designing new recipes by combining various ingredients.

The generated hyperlinks should have several desirable properties. In particular, they need to avoid being exact replicas of existing hyperlinks, while they should follow a distribution similar to those in the observed hypergraph. More specifically, the new hypergraph formed by these generated hyperlinks should preserve key characteristics of the observed hypergraph, such as node degree heterogeneity and hyperlink sparsity.
In the following section, we introduce our proposed approach, termed Denoising Diffused Embeddings, designed to address these requirements.



\subsection{Denoising Diffused Embeddings (DDE)}

Denoising Diffused Embeddings (DDE) is a generative modeling architecture for hypergraphs. The name reflects its two core components: first, we connect the hyperlink space to a latent embedding space through a conditional hyperlink likelihood; second, we reconstruct the latent space by training a diffusion model in which the latent embeddings are diffused through a stochastic noise process and then denoised to learn the underlying data-generating distribution.
Given an observed hypergraph $\cH([n],\cE_m)$, DDE outputs a synthetic hypergraph $\tilde{\cH}([n], \tilde{\cE}_{\tilde{m}} )$, where $\tilde{\cE}_{\tilde{m}} = \{\tilde{e}_1,\ldots,\tilde{e}_{\tilde{m}}\}$ represents the multiset of newly generated hyperlinks and $\tilde{m}$ denotes the number of these hyperlinks. In this section, we outline the components of DDE.

Assume that each node $i$ is associated with a latent embedding $\bz_i$ and a node degree parameter $\alpha_i\in\RR$. A hyperlink $e$ is generated based on its associated hyperlink embedding $\bx$, and its distribution is determined by a conditional likelihood model $\PP_{\cH([n],\{E\})|\bx,\cZ_n,\balpha_n}$, where $\cZ_n=\{\bz_1,\ldots,\bz_n\}$, $\balpha_n=(\alpha_1,\ldots,\alpha_n)^{\top}$, and $E$ denotes a random hyperlink.  The model $\PP_{\cH([n], \{E\})|\bx, \cZ_n, \balpha_n}$ is a probability measure defined over $\cP([n]) = \{e : e \subseteq [n]\}$, i.e., the collection of all subsets of $[n]$.
The difference between $\cZ_n$ and $\balpha_n$ is that $\cZ_n$ affects the likelihood through its interaction with $\bx$, while $\balpha_n$ affects the likelihood without interacting with $\bx$. 
For example, in the symptom co-occurrence hypergraph, each node embedding can be viewed as latent characteristics of the corresponding symptom, and the hyperlink embedding can be viewed as latent characteristics of a patient. The node degree parameters represent the general popularity of the symptoms. Whether a patient possesses certain symptoms is determined by its latent interactions with the symptom characteristics via $\cZ_n$ and $\bx$, as well as the general popularity of the symptoms. The interpretation of the likelihood model  depends on the specific model choice, and our framework permits the flexibility to choose the likelihood model that accommodates practical contexts. Specifically, DDE allows the embeddings to reside in general continuous latent spaces, including Euclidean and hyperbolic spaces, and supports different hyperlink likelihood models defined on these spaces (see Section \ref{sec:method} for more details).

Let $\cX_{m} =\{\bx_1,\ldots,\bx_m\}$ collects the hyperlink embeddings corresponding to the observed hypergraph  $\cH([n],\cE_m)$. 
Consider a \emph{random} hypergraph $\cH_{m,n} := \cH([n],\{E_1,\ldots,E_m\})$, from which $\cH([n],\cE_m)$ is a single realization. Let $\PP_{\cH_{m,n}|\cX_m,\cZ_n, \balpha_n }$ denote its distribution given the latent embeddings and node degree parameters.  This hypergraph distribution is then determined by
    $
       \PP_{\cH_{m,n}| \cX_m,\cZ_n,\balpha_n } = \prod_{j=1}^m \PP_{\cH([n], \{E_j\})| \bx_j,\cZ_n,\balpha_n}$.
Note that this product form implicitly imposes conditional independence among the hyperlinks given the latent embeddings and degree parameters. Similar conditional independence assumptions for hyperlink generation have been widely adopted in hypergraph analysis 
\citep[e.g.,][]{ke2019community,yuan2023high,zhen2023community}.
\begin{figure}
    \centering
    \includegraphics[width=0.97\textwidth]{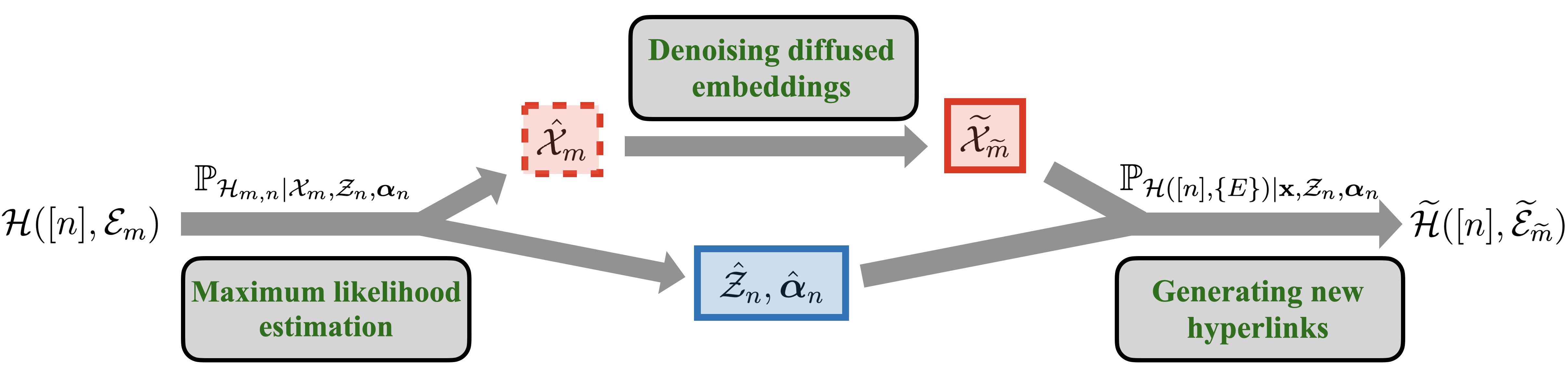}
    \caption{Denoising Diffused Embeddings.}
    \label{fig:dde_procedure}
\end{figure}

For a predetermined hypergraph model $\PP_{\cH_{m,n}| \cX_m,\cZ_n,\balpha_n }
$ and an observed hypergraph $\cH([n],\cE_m)$, DDE first embeds the observed discrete hyperlinks into the continuous embedding space via maximum likelihood estimation, and obtains estimators for the hyperlink embeddings, node embeddings, and degree parameters, denoted as $\hat{\cX}_m = \{\hat{\bx}_1,\ldots, \hat{\bx}_m\}$, $\hat{\cZ}_n = \{\hat{\bz}_1,\ldots,\hat{\bz}_n\}$, and $\hat{\balpha}_n = (\hat{\alpha}_1,\ldots,\hat{\alpha}_n)^{\top}$, respectively. Next, DDE trains a score-based diffusion model \citepalias{song2020score}, a class of nonparametric generative models on continuous spaces, on $\hat{\cX}_m$ to reconstruct the  hyperlink embedding space. To generate $\tilde{m}$ synthetic hyperlinks, DDE samples a multiset of $\tilde{m}$ new hyperlink embeddings, $\tilde{\cX}_{\tilde{m}} = \{\tilde{\bx}_1,\ldots,\tilde{\bx}_{\tilde{m}}\}$ from the diffusion model in the embedding space, and then generates a new hypergraph based on the conditional model $\prod_{j=1}^{\tilde{m}}\PP_{\cH([n],\{\tilde{E}_j\})|\tilde{\bx}_j,\hat{\cZ}_n,\hat{\balpha}_n}$, combining $\tilde{\cX}_{\tilde{m}}$, $\hat{\cZ}_n$, and $\hat{\balpha}_n$.
A graphical summary of DDE is in Figure \ref{fig:dde_procedure} and a detailed description of DDE is deferred to Section \ref{sec:method}.

\subsection{Related work}

With the rise of hypergraph data in the big data era, the modeling and analysis of hypergraph structures have garnered significant attention. \cite{ke2019community} highlighted that previous approaches projecting hypergraphs onto weighted graphs \citep{ghoshdastidar2015provable,kumar1812hypergraph,lee2020robust} often lead to information loss, yielding suboptimal community detection results.  \cite{jo2021edge} proposed a dual hypergraph transformation that connects graph edges to hypergraph nodes and applied message-passing techniques \citep{gilmer2017neural} for edge representation learning in graphs.  With regard to direct modeling of hypergraphs,  \cite{ke2019community} generalized the hypergraph planted partition model in \cite{ghoshdastidar2017consistency} and proposed a tensor-based algorithm for community detection. 
 \cite{yuan2023high} and \cite{zhen2023community}  modeled hypergraphs in latent spaces, and \cite{yuan2022testing} studied statistical inference for hypergraph  objectives.  
Despite this progress, these methods heavily rely on uniform restrictions of hyperlink orders and cannot handle the multiplicity of hyperlinks, i.e., multiple occurrences of the same hyperlink within a hypergraph. To overcome these limitations, \cite{wu2024general} proposed a latent embedding framework that supports varying hyperlink orders and repeated hyperlinks. This approach allows individual embeddings for both nodes and hyperlinks in the hypergraph, aligning well with the framework presented in this paper. 
While many areas of hypergraph analysis have been extensively studied, efficient and interpretable generative models for producing synthetic hyperlinks remain underexplored. 

Generative models are designed to produce data that mimics the features and structures of real-world observed data. Among these, diffusion models (\citealp{sohl2015deep}, \citetalias{song2020score}, \citealp{ho2020denoising}) have recently revolutionized applications such as image and audio generation, outperforming previous state-of-the-art methods like generative adversarial networks \citep[GAN,][]{creswell2018generative,goodfellow2020generative} and variational autoencoders \citep[VAE,][]{kingma2013auto,kingma2019introduction}. For an overview of recent progress on diffusion models, see \cite{chen2024opportunities}. Despite numerous methodological innovations, most diffusion models operate in continuous spaces. However, the discrete nature of hyperlinks requires generative models tailored to discrete spaces.  
Efforts to extend diffusion models to general discrete data include \cite{austin2021structured} and \cite{loudiscrete}, which operate directly on high-dimensional discrete variables and do not account for the unique characteristics and low-rank structure of hypergraphs \citep{thibeault2024low}. 
Consequently, these methods may exhibit slow convergence in training and sampling, yield suboptimal performance, and lack interpretability in hypergraph generation tasks. Existing graph generative models (\citealp{you2018graphrnn}, \citetalias{chen2023efficient}) focus on generating new discrete adjacency matrices, which are unsuitable for hyperlink generation because hyperlinks are discrete sets that cannot be represented by adjacency matrices.


In light of the growing demand for generating high-fidelity, interpretable hyperlinks with efficient training and sampling,
this work introduces DDE to address this gap by leveraging a  likelihood model which links the discrete hyperlinks to a continuous latent embedding space and improves interpretability, and a flexible while efficient diffusion model in the continuous latent  embedding space for fidelity. In training, DDE projects the high-dimensional hyperlinks onto a continuous, low-dimensional embedding space by training the hyperlink likelihood model, and then reconstruct that embedding space by training the diffusion model. Sampling is performed by first sampling embeddings from the latent diffusion model and then generating hyperlinks using the trained likelihood model and the generated hyperlink embeddings.
The idea of incorporating latent space low-dimensional structures and diffusion models has also been explored in recent studies.  With different training strategies, \cite{rombach2022high} combined diffusion models with autoencoder architectures to operate in the latent semantic spaces for image generation, while the latent score-based generative model \citep{vahdat2021score} functioned as a variant of VAE \citep{kingma2019introduction} with a diffusion model prior.
However, these approaches do not consider the unique characteristics of hypergraphs and lack interpretability, as the generated data are produced purely as black-box outputs. In contrast, DDE incorporates a likelihood-based modeling component that preserves key hypergraph properties, such as node degree heterogeneity and hyperlink sparsity, facilitates the study of model identifiability, and balances interpretability and fidelity in the generated data. We provide a more detailed comparison between VAE and DDE in Section A of the supplementary material.
When data resides on an unknown low-dimensional linear subspace, \citetalias{chen2023score} adapted the score network architecture to align with this structure and theoretically demonstrated that such modification allows diffusion models to circumvent the curse of ambient dimensionality.
In Section \ref{sec:theory}, we show that DDE provides a general solution to overcoming the curse of ambient dimensionality in high-dimensional hypergraphs. Additionally, we present a theoretical comparison between the approach in \citetalias{chen2023score} and DDE in Section C of the supplementary material.










\subsection{Contribution}

The contribution of this work can be summarized into two main aspects. Methodologically, we propose a general generative modeling architecture, DDE, specifically designed for hypergraphs. DDE leverages the unique properties of hypergraphs through likelihood modeling and incorporates score-based diffusion models to reconstruct the hyperlink embedding space,  which significantly improves training and sampling efficiency while preserving the fidelity of the generated data. Meanwhile, it provides interpretability during the generation of discrete and highly structured hypergraph data through the likelihood model, an interpretability that is lacking in many black-box generative models. We demonstrate the advantages of DDE over competing approaches using both simulated and real-world data.

Theoretically, we investigate the generative performance of DDE under both ideal and practical conditions.  In scenarios of perfect embedding estimation, where the node and hyperlink embeddings generating the observed hypergraph are accurately recovered, DDE fully circumvents the curse of ambient data dimensionality. Specifically, while a subset of $[n]$ can be encoded as a vector in $\{0,1\}^{n}$, we show that DDE reduces the task of generating $n$-dimensional discrete vectors to that of generating $K$-dimensional embeddings, where $K$, the latent space dimension, can be substantially smaller than $n$. Furthermore, we study how the embedding estimation procedure impacts the generative performance, highlighting the roles of dimensionality, node degree heterogeneity, and hyperlink sparsity of the hypergraph.  These technical
results not only advance theoretical understanding but also provide practical guidance for applying DDE effectively in hypergraph generative modeling, clarifying the conditions under which it performs optimally.

\subsection{Organization of the paper and notations}

The remainder of this article is organized as follows. Section \ref{sec:method} introduces the DDE framework. Section \ref{sec:theory} analyzes the generative performance of DDE. 
Section \ref{sec:simu} provides numerical comparisons of DDE with competing approaches. Section \ref{sec:realdata} applies DDE to the MIMIC-III dataset, generating new patient profiles and uncovering interesting discoveries. 
Finally, Section \ref{sec:discussion}
concludes the paper with a discussion. The supplementary material includes proofs of the main results, technical lemmas with their proofs, and additional numerical results.

For any $a,b\in\RR$, we denote their maximum and minimum by $a\vee b = \max(a,b)$ and $a\wedge b = \min(a,b)$, respectively. Given any two sequences $(a_{m,n}), (b_{m,n})$ valued in $\RR$, we say $a_{m,n} \lesssim b_{m,n}$ (or $a_{m,n} = O(b_{m,n})$,  $b_{m,n}\gtrsim a_{m,n}$) if there exists a universal constant $C > 0$ such that $|a_{m,n}| \le C|b_{m,n}|$ as $m,n\to\infty$. We say $a_{m,n}\asymp b_{m,n} $ if $a_{m,n}\lesssim b_{m,n}$ and $a_{m,n}\gtrsim b_{m,n}$. We write $a_{m,n}\ll b_{m,n}$ (or $a_{m,n}=o(b_{m,n})$) if $\lim_{m,n\to\infty} a_{m,n}/b_{m,n} = 0 $ and  $a_{m,n}\gg b_{m,n}$ (or $a_{m,n}=\Omega(b_{m,n})$) if $\lim_{m,n\to\infty} a_{m,n}/b_{m,n} = \infty $. We use $O_p(\cdot)$, $\Omega_p(\cdot)$, and $o_p(\cdot)$ as the stochastic versions of $O(\cdot)$, $\Omega(\cdot)$, and $o(\cdot)$ for convergence in probability.  For any positive integer $a$, we use $[a]$ to denote the set $\{1, 2, \ldots, a\}$. For any column vector $\ba$ and matrix $\bA$, we use $\ba^{\top}$ and $\bA ^ {\top}$ to denote their transposes, respectively. For an event $\mathfrak{E}$, the indicator function $\mathbf{1}_{\mathfrak{E}}=1$ if $\mathfrak{E}$ occurs and $0$ otherwise.  For a vector $\ba=(a_1,\cdots,a_n)^{\top}$, we use $\|\ba\|_2$ to denote the standard Euclidean norm and $\|\ba\|_{\infty} = \max_i|a_i|$ to denote the infinity/maximum/supreme norm. For a matrix $\bA=(a_{ji})_{m\times n}$, we use $\|\bA\|_2$. $\|\bA\|_{\max} = \max_{j\in[m],i\in[n]}|a_{ji}|$, and $\|\bA\|_{2\to\infty} = \max_{j\in[m]}\|\bA_{j,\cdot}\|_2 $ to denote its spectrum, supreme, and two-to-infinity norms, respectively. For a square matrix $\bA$, we use $\lambda_{\max}(\bA)$ and $\lambda_{\min}(\bA)$ to denote its maximal and minimal eigenvalues, respectively. For probability measures $\PP_1$ and $\PP_2$ defined on a measurable space $(\Omega,\cF)$, we denote their total variation distance as $\mathrm{d}_{\mathrm{TV}}(\PP_1,\PP_2) = \sup_{A\in\cF}|\PP_1(A) - \PP_{2}(A)|$ and KL (Kullback-Leibler) divergence as $\mathrm{d}_{\mathrm{KL}}(\PP_1~\|~\PP_2) = \int_{a\in\Omega} \log\big\{{\PP_1(da)}/{\PP_2(da)}\big\}\PP_{1}(da)$, respectively, where ${\PP_1(da)}/{\PP_2(da)}$ is the Radon-Nikodym derivative of $\PP_1$ with respect to $\PP_2$.

\section{Methodology}\label{sec:method}

In this section, we provide a detailed introduction to DDE. Algorithm \ref{alg:dde} illustrates the process by which DDE generates $\tilde{m}$ new hyperlinks from an observed hypergraph $\cH([n],\cE_m)$.
There are two modeling components in Algorithm \ref{alg:dde}. The first is the conditional hypergraph likelihood model, which builds the connection between the observed hypergraph and the latent embedding space. The second is a score-based diffusion model that is to be trained on the latent space and to sample new hyperlink embeddings from. We discuss these two modeling components and their implementation in Algorithm \ref{alg:dde} in the following.

\begin{algorithm}[t]
    \caption{Denoising Diffused Embeddings}
    \label{alg:dde}
    \KwIn{Hypergraph $\cH([n],\cE_m)$; Number of new hyperlinks $\tilde{m}$ ;  Conditional hypergraph likelihood model $\PP_{\cH_{m,n}|\cX_m,\cZ_n,\balpha_n} =  \prod_{j=1}^m \PP_{\cH([n],E_j) |\bx_j,\cZ_n,\balpha_n}$; Diffusion model architecture $\cM$. }  
    
\textbf{Training:} \textit{(i) Maximum likelihood estimation:} via $\PP_{\cH_{m,n}\mid \cX_m,\cZ_n,\balpha_n}$ and $\cH([n],\cE_m)$ to obtain estimated node embeddings $\hat{\cZ}_n=\{\hat{\bz}_1,\ldots,\hat{\bz}_n\}$, degree parameters $\hat{\balpha}_n=(\hat{\alpha}_1, \ldots,\hat{\alpha}_n)^{\top}$, and embeddings of observed hyperlinks $\hat{\cX}_m=\{\hat{\bx}_1,\ldots,\hat{\bx}_m\}$.
  \textit{(ii) Denoising diffused embeddings:} Train $\cM$ on $\hat{\cX}_m$ to obtain $\hat{\cM}$. \\

\textbf{Sampling:}  Independently for each $j\in[\tilde m]$, sample $\tilde{\bx}_{j}$ from $\hat{\cM}$ and then sample $\tilde e_j \sim \PP_{\cH([n],E)\mid \bx=\tilde{\bx}_j,\;\cZ_n=\hat{\cZ}_n,\;\balpha_n=\hat{\balpha}_n}$. Set $\tilde{\cE}_{\tilde m}=\{\tilde e_1,\ldots,\tilde e_{\tilde m}\}$. \\

    \KwOut{$\tilde{\cH}([n], \tilde{\cE}_{\tilde{m}})$.}
\end{algorithm}


For the conditional hypergraph likelihood model,
consider the random hypergraph $\cH_{m,n} = \cH([n],\{E_1, \ldots, E_m\})$, with each $E_j$ being a random subset of $[n]$.
The observed hypergraph $\cH([n],\{e_1,\ldots, e_m\})$ is one realization of $\cH([n],\{E_1,\ldots, E_m\})$ valued on $\{\cP([n])\}^{m}$. 
 The conditional hypergraph model then specifies 
\beq\label{eq:hyper_model}
\begin{aligned}
    & \PP_{\cH_{m,n}|\cX_m,\cZ_n,\balpha_n}\big\{ \cH([n],\{E_1,\ldots, E_m\}) = \cH([n],\{e_1,\ldots,e_m\}) \big\} \\ = &
   \prod_{j=1}^m\PP_{\cH([n],\{E_j\})|\bx_j,\cZ_n,\balpha_n}(E_j = {e_j}), ~~
   \text{for any }{e_1,\ldots,e_m}\in\cP([n]),
\end{aligned}
\eeq
through the hyperlink generating model $\PP_{\cH([n],\{E\})|\bx\,\cZ_n,\balpha_n}$. 
Below are three examples of such embedding models.

\begin{exmp}[Linear generation in Euclidean spaces \citep{wu2024general}]\label{exmp:linear}
    Consider $\bx$ valued in $\RR^K$ and $\cZ_n\subset\RR^{K}$, where $K$ is the latent Euclidean space dimension. Let $\sigma(a):= 1/ (1+ \exp(-a) )$ for $a\in\RR$ and $p_i(\bx) = \sigma(\bx^{\top}\bz_i + \alpha_i)$. \cite{wu2024general} models the conditional probability as $\PP_{\cH([n],\{E\})|\bx,\cZ_n,\balpha_n}(E = {e}) = \prod_{i\in e}p_i(\bx)\prod_{i\in[n]\setminus e}\big\{ 1 - p_i(\bx)\big\}$ for $e\in\cP([n])$ in \eqref{eq:hyper_model}.  As discussed in \cite{wu2024general}, this linear embedding approach effectively models the probability of any subset of nodes forming a hyperlink by considering the latent positions of the hyperlink and node embeddings. 
\end{exmp}

\begin{exmp}[Distance-based generation on general manifolds]\label{exmp:hyperbolic} Consider $\bx$ valued on $\HH$ and $\cZ_n\subset \HH$, where $\HH$ is a general manifold. 
Let $d_{\HH}(\bx, \bz_i)$ be the distance between $\bx$ and $\bz_i$ on $\HH$. Consider $\PP_{\cH([n],\{E\})|\bx,\cZ_n,\balpha_n}(E = {e}) = \prod_{i\in e}p'_{i}(\bx)\prod_{i\in[n]\setminus e}\big\{ 1 - p'_i(\bx)\big\}$, where $p'_{i} = g(-d_{\HH}(\bx,\bz_i),\alpha_i)$ for some function $g:\RR\times \RR\to[0,1]$ non-decreasing monotone in both its arguments. This embedding approach has been shown to effectively capture hierarchical geometry among graphical network nodes in hyperbolic latent spaces \citep{li2023statistical}. For diffusion models on general manifolds, one can refer to the work in \cite{huang2022riemannian}.
\end{exmp}

\begin{exmp}[Deterministic generation]\label{exmp:deterministic}
           Consider a deterministic function $f:\cX\times \cZ^n\times \RR^n \to \cP([n])$, where $\cX$ denotes the hyperlink embedding space and $\cZ$ denotes the node embedding space with $\cZ^n$ being the product of $n$ such spaces. Let $\PP_{\cH([n],\{E\})|\bx,\cZ_n,\balpha_n} = \delta_{f(\bx,\cZ_n,\balpha_n)}$ be a Dirac measure on $\cP([n])$, concentrated at $f(\bx,\cZ_n,\balpha_n)$. Under this generation rule, hyperlink generation is deterministic once the embeddings $\bx$, $\cZ_n$, and $\balpha_n$ are given. Such deterministic generation processes conditioned on the latent embeddings have been considered in \cite{li2023spectral} for community detection in block models. 
\end{exmp}

 Different embedding models serve different interpretative purposes and apply to different scenarios. Fixing the likelihood model, Step (i) in the training stage of Algorithm \ref{alg:dde} uses maximum likelihood estimation on $\cH([n],\cE_m)$ to obtain estimated embeddings $\hat{\cX}_m$, $\hat{\cZ}_n$ and degree parameters $\hat{\balpha}_n$.  
Step (ii) in the training stage trains a score-based diffusion model on the estimated embeddings $\hat{\cX}_m$ to reconstruct the hyperlink embedding space. In sampling, DDE first generates synthetic hyperlink embeddings $\tilde{\bx}_j$ and then generates hyperlinks from the likelihood model $\PP_{\cH([n],E)\mid \bx=\tilde{\bx}_j,\;\cZ_n=\hat{\cZ}_n,\;\balpha_n=\hat{\balpha}_n}$, plugging in the generated hyperlink embeddings and the estimated node embeddings and degree parameters. In the following, we further illustrate the procedure using the linear embedding model in Example~\ref{exmp:linear} for the hyperlink likelihood model and score-based generative modeling through stochastic differential equations \citepalias{song2020score} for the diffusion model architecture.

\subsection{Embedding hyperlinks}\label{sec:method:embedding}

Given the conditional hypergraph likelihood model, we need to estimate the latent embeddings and degree parameters from the observed hypergraph $\cH([n],\cE_m)$. Consider Example~\ref{exmp:linear}. We first address the identifiability issue of the model parameters. Note that for any $\bX$ valued in $\RR^K$, $\bmu_\bX \in \RR^K$ and full-rank $\bA \in \RR^{K \times K}$, by letting
\beq\label{eq:trans_A}
   {\bX}' = \bA^{\top}(\bX - \bmu_\bX),~ {\bz}_i' = \bA^{-1}\bz_i, ~\text{and}~~\alpha_i' = \alpha_i + \bmu_{\bX}^{\top}\bz_i,~ \text{for} ~ i\in[n],  
\eeq
we have $\sigma(\bX^{\top}{\bz_i} + \alpha_i) = \sigma({\bX'}^{\top}\bz_i' + \alpha_i')$ for all $i\in[n]$. This indicates that $\PP_{\bX}$ and $\cZ_n$ are not identifiable due to a shift in $\PP_{\bX}$ and an invertible linear transformation between $\bX$ and $\{\bz_1,\ldots,\bz_n\}$, as these transformations do not change the hyperlink distribution. To address this issue, we consider the following identifiability conditions on the hyperlink embedding distribution and the node embeddings:
    $\text{(C1)}~ \EE_{\PP_{\bX}}[\bX] = \mathbf{0}_K$, $\text{(C2)}~ \EE_{\PP_{\bX}}[\bX\bX^{\top}]=\lim_{n\to\infty}\bZ_n^{\top}\bZ_n/n~\text{and both of them are diagonal,}$ 
where $\bZ_n = (\bz_1,\ldots,\bz_n)^{\top}\in\RR^{n\times K}$.

\begin{rem}
    The identifiability conditions (C1) and (C2) are for controlling the above shift issue in $\PP_{\bX}$ and the invertible linear transformation between $\bX$ and $\{\bz_1,\ldots,\bz_n\}$, respectively, to guarantee the  estimatibility of $\PP_{\bX}$, $\{\bz_1,\ldots,\bz_n\}$, and $\balpha_n$.  Such identifiability conditions have been widely considered in the literature; see, for example, \cite{bai2012statistical}, \cite{wang2022maximum} and \cite{wu2024general}. 
    Another commonly used set of conditions is that $\EE_{\PP_{\bX}} = \mathbf{0}_K$, $\EE_{\PP_{\bX}}[\bX\bX^{\top}] = \bI_K$, and $\lim\limits_{n\to\infty}\bZ_n^{\top}\bZ_n/n$ is diagonal. These conditions can be transformed to (C1) and (C2) by left-multiplying $(\lim\limits_{n\to\infty}\bZ_n^{\top}\bZ_n/n)^{1/4}$ on $\bX$ and $(\lim\limits_{n\to\infty}\bZ_n^{\top}\bZ_n/n)^{-1/4}$ on each of $\{\bz_1,\ldots,\bz_n\}$. We choose conditions (C1) and (C2) for technical purposes in the theoretical analysis. Similar theoretical results can also be derived for other identifiability conditions that guarantee the estimatibility of $\PP_{\bX}$, $\{\bz_1,\ldots,\bz_n\}$, and $\balpha_n$.   
\end{rem}

\begin{rem}
    In general, if the true $\PP_{\bX}$ and $\cZ_n$ do not satisfy these constraints, a transformation of $\PP_{\bX}$, $\cZ_n$, and $\balpha_n$ can be made so that the constraints are met and the hyperlink distribution remains unchanged. In particular, let $\balpha_n' = \balpha_n + \bZ_n\cdot\EE_{\PP_{\bX}}[\bX]$. Let
  $\cV = \diag( \rho_1^2,\cdots,\rho_K^2 )$ with $\rho_1^2>\rho_2^2>\cdots >\rho_K^2$ being the eigenvalues of $(\lim\limits_{n\to\infty}\bZ_n^{\top}\bZ_n/n)^{1/2}\EE_{\PP_{\bX}}\big[ (\bX - \EE_{\PP_{\bX}}[\bX] )(\bX - \EE_{\PP_{\bX}}[\bX] )^{\top} \big](\lim\limits_{n\to\infty}\bZ_n^{\top}\bZ_n/n)^{1/2}$, and let $\bGamma$ be the $K\times K$ matrix whose columns collect the corresponding eigenvectors.   Let $\bX' = \bG^{\top}(\bX - \EE_{\PP_{\bX}}[\bX])$ and $\bZ_n' = \bZ_n(\bG^{-1})^{\top}$, where $\bG = \big(\lim\limits_{n\to\infty}\bZ^{\top}_n\bZ_n/n \big)^{1/2}\bGamma\cV^{-\frac{1}{4}}$. Then $\PP_{\bX'}$, $\bZ_n'$ and $\balpha_n'$ will satisfy identifiability conditions (C1) and (C2) while keeping the hyperlink distribution unchanged. 
Another interpretation of the transformation is that we fix $\bmu_{\bX} = \EE_{\PP_{\bX}}[\bX]$ and  
$\bA = \bG$ in \eqref{eq:trans_A}. 
\end{rem}


After resolving the identifiability issues arising from the shift and linear transformation, 
\((\PP_{\bX},\bZ_n,\balpha_n)\) are identifiable up to coordinate-wise sign changes. 
This remaining identifiability issue can be easily fixed by imposing a sign convention, such as fixing the signs of coordinates of a selected reference node embedding.
Throughout this article, we assume that $\{\bz_1,\ldots,\bz_n\}$ are bounded and $\PP_{\bX}$ is supported on a bounded set. Define $\bar{\alpha}_{m,n} = \sum_{i=1}^n\alpha_i/n$ to be the average node degree parameter, where the subscripts $m,n$ are used to characterize the asymptotic regime as $m,n\to\infty$. We assume that $|\alpha_i - \bar{\alpha}_{m,n}|$ is bounded for all $i\in[n]$ and $\bar{\alpha}_{m,n}$ is upper bounded but can diverge to $-\infty$ as $m,n\to\infty$.  The expected order of a hyperlink $E$ is then 
\[
  \EE[\sum_{i=1}^n\mathbf{1}_{\{i\in E\}}] = \EE_{\PP_{\bX}}(\EE[\sum_{i=1}^n\mathbf{1}_{\{i\in E\}}|\bX] )=\EE_{\PP_{\bX}}\Big[\sum_{i\in[n]}\frac{\exp(\bX^{\top}\bz_i + \alpha_i)}{1 + \exp(\bX^{\top}\bz_i + \alpha_i)}\Big]  \asymp n\exp(\bar{\alpha}_{m,n}),
\]
as $m,n\to\infty$, which can be much smaller than $n$ as $\bar{\alpha}_{m,n}\to-\infty$. This is consistent with the observation in hypergraph data that the orders of hyperlinks are usually much smaller compared with the total number of nodes. Theoretical analyses in Section \ref{sec:theory} will further illustrate how the rate of $\exp(\bar{\alpha}_{m,n})$, i.e., the hyperlink sparsity, affects the performance of DDE. 
Accordingly, 
for an observed hypergraph $\cH([n],\{e_1,\ldots,e_m\})$,  let $\bX_m = (\bx_1,\ldots,\bx_m)^{\top}\in\RR^{m\times K}$, and
consider the log-likelihood function 
$$
\begin{aligned}
 \cL(\cX_m,\cZ_n,\balpha_n) := & \cL(\cH([n],\{e_1,\ldots,e_m\}) |\cX_m,\cZ_n,\balpha_n) \\
 = & \sum_{j=1}^m\sum_{i=1}^n\big\{ \mathbf{1}_{i\in e_j}(\bx_j^{\top} \bz_i+\alpha_i) - \log\big( 1+ \exp(\bx^{\top}_j\bz_i+\alpha_i )   \big)    \big\}  
\end{aligned}
$$ 
We then consider the following constrained maximum likelihood estimator
\beq\label{eq:cmle}
\begin{aligned}
   & \max_{\cX_m,\cZ_n,\balpha_n} \cL(\cX_m,\cZ_n,\balpha_n) \\ ~~\text{s.t.}~~ & \frac{1}{n}\bZ^{\top}_n\bZ_n= \frac{1}{m}\bX^{\top}_m\bX_m,      ~\bZ_n^{\top}\bZ_n  \text{ and } \bX^{\top}_m\bX_m \text{ are diagonal}, \bX_{m}^{\top}\mathbf{1}_m = \mathbf{0}_K, \\ &
   \max\{\|\bZ_n\|_{\max},\|\bX_m\|_{\max},\|\balpha_n -\frac{\sum_{i=1}^n\alpha_i}{n}\cdot\mathbf{1}_n\|_{\infty}\} \le C , \\ &
  \text{ and } -C_{m,n} \le n^{-1}\sum_{i=1}^n\alpha_i \le -C'C_{m,n},
\end{aligned}
\eeq
for some constants $C>0$, $C'\in(0,1)$, and a boundary parameter $C_{m,n}>0$ that can slowly diverge as $m,n\to\infty$ to control the hyperlink sparsity. We suggest setting $C_{m,n} = -C''\log\{\sum_{j=1}^m|e_j|/(mn)\}$ for some $C''>1$ following Proposition 3.2 in \cite{wu2024general}. 
We defer the algorithms for computing \eqref{eq:cmle} to Section E.1.1 of the supplementary material.
The constraints in \eqref{eq:cmle} are from the identifiability conditions (C1) and (C2) by noticing that $\cX_m=\{\bx_1,\ldots,\bx_m\}$ are $m$ realizations of $\PP_{\bX}$. Let $\hat{\cX}_m = \{\hat{\bx}_1,\ldots,\hat{\bx}_m\}$, $\hat{\cZ}_n  = \{\hat{\bz}_1,\ldots, \hat{\bz}_n \}$ and $\hat{\balpha}_n = (\hat{\alpha}_1,\ldots,\hat{\alpha}_n)^{\top}$ be the solution to \eqref{eq:cmle}. We have now completed Step (i) of training in Algorithm \ref{alg:dde} under the linear embedding approach.

\begin{rem}
   The choice of embedding space dimension $K$ is important for the algorithm, as it determines the specific form of the likelihood model and the parameter space. Under this framework, the latent space dimension can be selected in a data driven way using existing methods such as cross-validation \citep{li2020network} or information criteria \citep{chen2022determining}. We provide a detailed discussion of the latent space dimension in Section D of  the supplementary material. Meanwhile, our empirical results show that DDE is robust to such misspecification, as in the synthetic medical record generation task in Section~\ref{sec:realdata}. 
\end{rem}

\subsection{Score-based diffusion models in the embedding space}\label{sec:score-based}

Next, we introduce the diffusion model architecture  (\citealp{sohl2015deep}, \citetalias{song2020score}, \citealp{ho2020denoising}) in the embedding space. In this work, we consider the framework of score-based generative modeling via stochastic differential equations (SDE) \citepalias{song2020score}, which gradually perturbs the embeddings into pure noise via a forward SDE and learns a backward SDE recovering the embedding distribution  via score matching (\citealp{hyvarinen2005estimation}, \citealp{vincent2011connection}, \citetalias{song2020score}).

Specifically, we first construct a continuous-time diffusion process $\{\bX_t\}_{t \in [0,T]}$ that is modeled as the solution to an It\^o SDE:
$d\bX_t = \bff(\bX_t, t)\,dt + g(t)\,d\bW_t$ with $\bX_0$ following the distribution of estimated hyperlink embeddings.
Here, $\bff(\cdot, t): \RR^K \to \RR^K$ is a vector-valued function representing the drift coefficient of $\bX_t$, $g(\cdot): \RR \to \RR$ is a scalar-valued diffusion coefficient, and $\bW_t$ is a standard Wiener process. This SDE has a unique strong solution as long as $\bff$ and $g$ are globally Lipschitz in both state and time \citep{oksendal2003stochastic}. Let $\PP_{\bX_t}$ denote the distribution of $\bX_t$. The functions $\bff$ and $g$ are chosen such that $\PP_{\bX_T}$ is close to a tractable distribution that is easy to sample from, which we take to be the standard Gaussian in this article.
The reverse of this process is given by the backward SDE: $
d\bX_t^{\leftarrow} = \left\{ \bff(\bX_t^{\leftarrow}, t) - g(t)^2 \nabla \log p_{T-t}(\bX_t^{\leftarrow}) \right\} dt + g(t) \, d\tilde{\bW}_t$ with $\bX_0^{\leftarrow} \sim \PP_{\bX_T}$,
where $\nabla \log p_t(\cdot)$ denotes the score function, i.e., the gradient of the log-density of $\bX_t$, and $\tilde{\bW}_t$ is another  Wiener process. The superscript ``$\leftarrow$'' distinguishes the backward process from the forward process. Under mild conditions,  $\{\bX_t^{\leftarrow}\}_{0 \le t < T}$ initialized from $\PP_{\bX_T}$ has the same distribution as the time-reversed forward process $\{\bX_{T - t}\}_{0 \le t < T}$ \citep{haussmann1986time}.

In practice, directly sampling from the backward SDE is infeasible, as both the score functions and the distribution of $\bX_T$ are unknown. To address this, we replace $\PP_{\bX_0^{\leftarrow}}$ in the backward SDE with a standard multivariate Gaussian distribution, uniformly discretize the process with a step size \( h > 0 \), and train neural networks to estimate the score functions. 
Specifically, let \( N = T/h \) denote the number of discretization steps. Let \( \bs_{\btheta}(\cdot, \cdot): \RR^K \times [0,T] \to \RR^K \) be the score network, a ReLU network parameterized by \( \btheta \) for score approximation \citep{chen2023score}. The parameter \( \btheta \) is learned via the denoising score matching objective \citep{vincent2011connection} defined on learned embeddings:
\begin{equation} \label{eq:score-estimate}
  \hat{\btheta} \in \argmin_{\btheta} \sum_{k=0}^{N} \lambda_k \cdot \frac{1}{m} \sum_{\bx_0 \in \hat{\cX}_m} \EE_{\bx_{kh}|\bx_0} \left\| \bs_{\btheta}(\bx_{kh}, kh) - \nabla_{\bx_{kh}} \log p_{kh}(\bx_{kh}|\bx_0) \right\|_2^2,
\end{equation}
where \( \lambda_k \approx 1/\EE\left\{ \left\| \nabla_{\bx_{kh}} \log p_{kh}(\bx_{kh}|\bx_0) \right\|_2^2 \right\} \) are positive weights for different time steps, and \( \bx_{kh}|\bx_0 \sim p_{kh}(\bx_{kh}|\bx_0) \) denotes the conditional distribution of \( \bX_{kh} \) in the forward process initialized at \( \bx_0 \). Step (ii) in the training stage of Algorithm \ref{alg:dde} is then completed.

In the sampling stage of Algorithm \ref{alg:dde},
to generate samples from the diffusion model, we simulate the following plug-in backward SDE using the exponential integrator scheme \citep{song2020denoising, zhang2022fast}. Specifically, for each \( k = 0, \ldots, N-1 \), and for \( t \in [kh, (k+1)h] \), we freeze the score network at time \( kh \), which yields:
\beq\label{eq:back-sde}
d\bX_t^{\leftarrow} = \left\{ \bff(\bX_t^{\leftarrow}, t) - g(t)^2 \bs_{\hat{\btheta}}(\bX_{kh}^{\leftarrow}, T-kh) \right\} dt + g(t) \, d\tilde{\bW}_t, ~t\in[kh,(k+1)h],
\eeq
where \( \hat{\btheta} \) is the score network parameter trained via \eqref{eq:score-estimate}. 
We simulate the backward SDE \eqref{eq:back-sde} \( \tilde{m} \) times with initial states \( \bX_0^{\leftarrow} \) drawn from a standard multivariate Gaussian, resulting in \( \tilde{\cX}_m = \{ \tilde{\bx}_1, \ldots, \tilde{\bx}_{\tilde{m}} \} \). Each of the $\tilde{m}$ hyperlinks are then generated from the likelihood model $\PP_{\cH([n],E)\mid \bx=\tilde{\bx}_j,\;\cZ_n=\hat{\cZ}_n,\;\balpha_n=\hat{\balpha}_n}$ for $j\in[\tilde{m}]$, with $\hat{\cZ}_n$ and $\hat{\balpha}_n$ from \eqref{eq:cmle}.

\begin{rem}
    Compared with existing work on generative modeling using latent space diffusion models \citep{vahdat2021score, rombach2022high}, the DDE framework has several advantages and  contributions. First, it respects the unique properties and structure of hypergraphs through conditional likelihood modeling, which also offers interpretability to the generated samples by adapting the likelihood to hypergraph generation tasks across domains. Second, it provides an efficient training and sampling scheme that avoids training deep models on high-dimensional parameter spaces and does not rely on variational approximations or iterative training between the decoder and latent space diffusion prior \citep{vahdat2021score}. Third, it admits theoretical understanding of the generated samples with connections to hypergraph properties, which is lacking in many existing approaches. 
\end{rem}

\begin{rem}
  As a general generative modeling framework, the latent space diffusion component in DDE can be replaced by other generative models that can be trained on estimated hyperlink embeddings and that permit sampling of high-fidelity embeddings. We consider score-based diffusion models for several reasons. First, they are more flexible than simple parametric models on the latent space, which can be overly restrictive for the generation task. Second, they are more sample-efficient than traditional nonparametric density estimators in moderately high dimensions \citep{silverman2018density}, which suffer severely from the curse of dimensionality. Finally, the strong theoretical justification for the empirical success of diffusion models with minimal assumptions \citep{chen2022sampling} makes them appealing in our application and facilitates the theoretical analysis in this work.
\end{rem}

\begin{rem}
   DDE learns the hyperlink distribution at the out-of-sample population level for the generation task, in contrast to the in-sample perspective in \cite{wu2024general}. Here, \emph{out-of-sample} means that generated hyperlinks are not identically distributed to the observed hyperlinks conditional on their latent hyperlink embeddings; instead, DDE learns the \emph{unconditional} distribution of hyperlinks and generates samples that approximate the broader population distribution rather than reproducing a particular conditional distribution in the observed hypergraph.
This design is practically important. For example, in the MIMIC-III medical record application, each embedding $\bx_j$ encodes latent patient characteristics and can be consistently estimated with access to new hyperlinks associated with that embedding, as shown in Theorem~\ref{thm:uniform_consis}. By learning the unconditional distribution and generating out-of-sample, DDE avoids directly sharing  individual hyperlink information. 
\end{rem}




\section{Theoretical  analysis}\label{sec:theory}


Our theoretical analysis in this section considers the following setup. The node embeddings $\cZ_n=\{\bz_1,\ldots,\bz_n\}$ and node degrees in $\balpha_n = (\alpha_1,\ldots,\alpha_n)^{\top}$ are 
treated as fixed parameters. The hyperlink embeddings $\cX_m = \{\bx_1,\ldots,\bx_m\}$ consist of $m$ realizations drawn from the hyperlink embedding distribution $\PP_{\bX}$. The observed hypergraph is then generated from $\cX_m$, $\cZ_n$, and $\balpha_n$, based on the conditional distribution $\PP_{\cH_{m,n}|\cX_m,\cZ_n,\balpha_n}$, which can be decomposed into independent hyperlink generation processes: $\PP_{\cH_{m,n}|\cX_m,\cZ_n,\balpha_n} = \prod_{j=1}^m \PP_{\cH([n],(E_j))|\bx_j,\cZ_n,\balpha_n}$. Our goal is to evaluate how ``close'' the distribution of the hyperlinks generated by DDE (Algorithm \ref{alg:dde}) is to the distribution of hyperlinks generated by the true model, i.e., by $\PP_{\bX}$ and $\PP_{\cH([n],\{E\})|\bX,\cZ_n,\balpha_n}$. 
We first present a general theorem on DDE  under the assumption that the true embeddings and parameters are available.
Let $\PP_{(E,\bX)}$ denote the joint distribution of a hyperlink $E$ and its associated hyperlink embedding $\bX$.
Let $\PP_{\tilde{\bX}}$ denote the marginal distribution of a hyperlink embedding $\tilde{\bX}$ sampled from diffusion models trained on $\{\bx_1,\ldots,\bx_m\}$. Additionally, let $\PP_{(\tilde{E},\tilde{\bX})}$ denote the joint distribution of $\tilde{\bX}$ and a hyperlink generated from this embedding along with $\cZ_n$ and $\balpha_n$. We establish the following lemma. 

\begin{lem}\label{thm:KL} $
      \mathrm{d}_{\mathrm{KL}}( \PP_{(E,\bX)} \|\PP_{(\tilde{E}, \tilde{\bX} )} ) = \mathrm{d}_{\mathrm{KL}}(\PP_{\bX}\|\PP_{\tilde{\bX}}) $. 
\end{lem}

Lemma \ref{thm:KL} demonstrates that DDE reduces
the generative error for  high-dimensional hyperlinks to the generative error for low-dimensional hyperlink embeddings under the ideal scenario where $\cX_m$, $\cZ_n$, and $\balpha_n$ are available.
In cases where the embeddings and node degrees must be estimated, we introduce the distributions $\PP'_{(\tilde{E},\tilde{\bX})}$ and $\PP'_{\tilde{\bX}}$. 

\begin{defn}\label{def:ptilde}
    Conditioned on $\cX_m$,
    we use $\PP'_{(\tilde{E},\tilde{\bX})}$ to denote a random measure on the joint distribution of a generated hyperlink embedding $\tilde{\bX}$ and the hyperlink $\tilde{E}$ generated from $\tilde{\bX}$, $\hat{\cZ}_n$ and $\hat{\balpha}_n$. More formally, this random measure should  be denoted by $\PP'_{(\tilde{E},\tilde{\bX})|\hat{\cX}_m,\hat{\cZ}_n,\hat{\balpha}_n}$, and here we use the notation $\PP'_{(\tilde{E},\tilde{\bX})}$  for simplicity. The distribution of this random measure depends on the $m$ realizations of hyperlink embeddings $\cX_{m}$.  The randomness in $\PP'_{(\tilde{E},\tilde{\bX})}$ is from the hypergraph generation process given $\cX_m$, $\cZ_n$ and $\balpha_n$, which introduces randomness in the estimated embeddings $\hat{\cX}_m$, $\hat{\cZ}_n$ and node degrees $\hat{\balpha}_n$, thus leading to randomness in the distribution of $\tilde{\bX}$ and $\tilde{E}$.  Similarly, $\PP'_{\tilde{\bX}}:= \PP'_{\tilde{\bX}|\hat{\cX}_m,\hat{\cZ}_n,\hat{\balpha}_n}$ is a random measure on $\tilde{\bX}$ conditioned on $\cX_m$. 
\end{defn}

 
The following theorem decomposes the distance between $\PP_{(E,\bX)}$ and $\PP'_{(\tilde{E},\tilde{\bX})}$.

\begin{thm}\label{thm:KL_noisy} The $\mathrm{KL}$-divergence between $\PP_{(E,\bX)}$ and $\PP'_{(\tilde{E},\tilde{\bX})}$ can be decomposed into three terms as follows:
\[
    \mathrm{d}_{\mathrm{KL}}(\PP_{(E,\bX)} \| \PP'_{(\tilde{E},\tilde{\bX})}) = \mathrm{error}_{(\bz,\alpha)\text{-}\hspace{0.1em}\mathrm{estimation}} + \mathrm{error}_{\PP_{\bX}\text{-}\hspace{0.1em}\mathrm{estimation}} + \mathrm{error}_{\mathrm{diffusion}}. 
\]
Specifically, the three terms take the following forms:
\begin{align*}
& \mathrm{error}_{(\bz,\alpha)\text{-}\hspace{0.1em}\mathrm{estimation}}  =  \EE_{\PP_{(E,\bX)}} \Bigg[ \log\Bigg\{ \frac{ \frac{d\PP_{\cH([n],\{E\})|\bX,\cZ_n,\balpha_n}}{d\mu(  \cH([n],\{E\}))} (E,\bX) }{\frac{d\PP_{\cH([n],(\tilde{E}))|\tilde{\bX},\hat{\cZ}_n,\hat{\balpha}_n}}{d\mu(  \cH([n],(\tilde{E})))} (E,\bX)  } \Bigg\} \Bigg], \\
& \mathrm{error}_{\PP_{\bX}\text{-}\hspace{0.1em}\mathrm{estimation}} =  \EE_{\PP_{\bX}}\Bigg[ \log \Bigg\{\frac{ \frac{d\PP_{\bX}}{d\mu(\bX)} (\bX)  }{ \frac{d\PP_{\hat{\cX}_m}}{d\mu(\hat{\bX})} (\bX) }\Bigg\}  \Bigg] + \EE_{\PP_{\bX}}\Bigg[ \log \Bigg\{\frac{ \frac{d\PP_{\hat{\cX}_m}}{d\mu(\hat{\bX})} (\bX)  }{ \frac{d\PP'_{\tilde{\bX}}}{d\mu(\tilde{\bX})} (\bX) }\Bigg\}  \Bigg] \\
       &~~~~~~~~~~~~~~~~~~~~~~~ - \EE_{\PP_{\hat{\cX}_m}}\Bigg[ \log \Bigg\{\frac{ \frac{d\PP_{\hat{\cX}_m}}{d\mu(\hat{\bX})} (\bX)  }{ \frac{d\PP'_{\tilde{\bX}}}{d\mu(\tilde{\bX})} (\bX) }\Bigg\}  \Bigg],  ~\text{and}\\
 &   \mathrm{error}_{\mathrm{diffusion}}    = \EE_{\PP_{\hat{\cX}_m}}\Bigg[ \log \Bigg\{\frac{ \frac{d\PP_{\hat{\cX}_m}}{d\mu(\hat{\bX})} (\bX)  }{ \frac{d\PP'_{\tilde{\bX}}}{d\mu(\tilde{\bX})} (\bX) }\Bigg\}  \Bigg],
\end{align*}
where $\PP_{\hat{\cX}_m}$ stands for the marginal distribution of each one of $\{\hat{\bx}_1,\ldots,\hat{\bx}_m\}$ given $\cX_m$ and we assume absolute continuity of the log ratios without loss of generality.  
\end{thm}

The error decomposition in Theorem \ref{thm:KL_noisy} applies to
general conditional likelihood model $\PP_{\cH([n],\{E\})|\bx,\cZ_n,\balpha_n}$ and diffusion model architecture. In the following subsections, we analyze each error term under the setup of Sections 
\ref{sec:method:embedding} and \ref{sec:score-based}.

\subsection{On $ \mathrm{error}_{(\bz,\alpha)\text{-}\hspace{0.1em}\mathrm{estimation}}$ and node embedding estimation}

The first term in Theorem \ref{thm:KL_noisy} is referred to as $\mathrm{error}_{(\bz,\alpha)\text{-}\hspace{0.1em}\mathrm{estimation}}$ since $\frac{d\PP_{\cH([n],\{E\})|\bX,\cZ_n,\balpha_n}}{d\mu(  \cH([n],\{E\}))} (E,\bX)$ and $\frac{d\PP_{\cH([n],(\tilde{E}))|\tilde{\bX},\hat{\cZ}_n,\hat{\balpha}_n}}{d\mu(  \cH([n],(\tilde{E})))} (E,\bX) $ only differ in $\hat{\cZ}_{n}$ and $\hat{\balpha}_n$ with the same plugged-in $(E,\bX)$ pairs. The second term, 
$\mathrm{error}_{\PP_{\bX}\text{-}\hspace{0.1em}\mathrm{estimation}}$, is so named because it primarily depends on the distance between $\PP_{\bX}$ and $\PP_{\hat{\cX}_m}$. The third term $\mathrm{error}_{\mathrm{diffusion}}$ quantifies the error in generating observations of $\PP_{\hat{\cX}_m}$ using diffusion models. We analyze the three terms in sequence.

Consider the linear embedding approach in Example \eqref{exmp:linear}, and that $\PP_{\bX}$, $\cZ_n$, and $\balpha_n$ satisfy the identifiability conditions (C1) and (C2).  We analyze $\mathrm{error}_{(\bz,\alpha)\text{-}\hspace{0.1em}\mathrm{estimation}}$ and $\mathrm{error}_{\PP_{\bX}\text{-}\hspace{0.1em}\mathrm{estimation}}$ under an asymptotic regime where both $m$ and $n$ go to infinity and consider the following conditions. 
 
\begin{cond}[Embedding assumptions]\label{cond:eigen-structure}
    The node embeddings and degrees satisfy $ \max\\\{\max_{i\in[n]}\|\bz_i\|_{\infty},\max_{i\in[n]}|\alpha_i - \sum_{i=1}^n\alpha_i/n| \}\le C$ for the constant $C$ in \eqref{eq:cmle} and also the support of $\PP_{\bX}$ satisfies $\supp(\PP_{\bX})\subset \{\bx\in\RR^K:\|\bx\|_{\infty}\le C\}$. For the $m$ realizations of $\PP_{\bX}$ collected in $\cX_{m}$, they satisfy that $\|m^{-1}\sum_{j=1}^m \bx_j\|_2 = O\big( \{(m\wedge n) \exp(\bar{\alpha}_{m,n})\}^{-{1}/{2}}  \big)$ and $ \big\|\bX^{\top}_m\bX_m/m  - \EE_{\PP_{\bX}}[\bX\bX^{\top}] \big\|_2 =O\big( \{(m\wedge n) \exp(\bar{\alpha}_{m,n})\}^{-1}  \big) $. The minimum eigenvalue of 
    $\EE_{\PP_{\bX}}[\bX\bX^{\top}]$  is lower bounded by a constant, and eigenvalues of $\EE_{\PP_{\bX}}[\bX\bX^{\top}]$
    are distinct and their gaps are lower bounded by a constant. 
\end{cond}

\begin{cond}[Hyperlink sparsity assumptions]\label{cond:node}
    The average node degree parameter $\bar{\alpha}_{m,n}$ satisfies that $\exp(\bar{\alpha}_{m,n})\gtrsim (m\vee n)^{\epsilon^*}/(m\wedge n)$ for some arbitrarily small $\epsilon^*>0$. 
\end{cond}


Condition \ref{cond:eigen-structure} assumes bounded embeddings and degree heterogeneity, and a reasonable convergence rate of the first and second moments of $\cX_{m}$ to their population counterparts. Notably, for sparser hypergraphs, the required rate of this convergence is less stringent. Condition \ref{cond:node} specifies the minimum expected hyperlink order requirement, which is $n(m\vee n)^{\epsilon^*}/(m\wedge n)$ for some small $\epsilon^*>0$. 
Under these conditions, we have the following Theorem \ref{thm:uniform_consis} which will be used in analyzing $\mathrm{error}_{(\bz,\alpha)\text{-}\hspace{0.1em}\mathrm{estimation}}$ and $\mathrm{error}_{\PP_{\bX}\text{-}\hspace{0.1em}\mathrm{estimation}}$.

\begin{thm}\label{thm:uniform_consis}
    Let ${\bx}_1,\ldots,{\bx}_m$ be $m$ realizations from $\PP_{\bX}$ and $\cH([n],\cE_m)$ be generated from $\cX_m=\{\bx_1,\ldots,\bx_m\}$, $\cZ_n=\{\bz_1,\ldots,\bz_n\}$ and $\balpha_n = (\alpha_1,\ldots,\alpha_n)^{\top}$. Let  $\hat{\cX}_m = \{\hat{\bx}_1,\ldots,\hat{\bx}_m\}$, $\hat{\cZ}_n = \{\hat{\bz}_1,\ldots,\hat{\bz}_n\}$ and $\hat{\balpha} =(\hat{\alpha}_1,\ldots,\hat{\alpha}_n)^{\top}$ be the solution to \eqref{eq:cmle}, where we set $C_{m,n} = -C''\log\{\sum_{j=1}^m|e_j|/(mn)\}$ for some $C''>1$. Suppose Conditions \ref{cond:eigen-structure} and \ref{cond:node} hold. Then for any $\epsilon>0$, as $m,n\to\infty$, we have 
    \[
    \begin{aligned}
& m^{-1}\sum_{j=1}^{m} \|\hat{\bx}_j - \bx_j\|_2^2
+
n^{-1} \sum_{i=1}^{n} \|\hat{\bz}_i - \bz_i\|_2^2
+
n^{-1} \|\hat{\balpha} - \balpha\|_2^2
=
O_p \big\{(m \wedge n)^{-1} \exp(-\bar{\alpha}_{m,n}) \big\},~\text{and} \\
    &  \max\big\{\max_{j\in[m]}\| \hat{\bx}_j - \bx_j \|_{\infty}, \max_{i\in[n]} \| \hat{\bz}_i - \bz_i \|_{\infty} ,\max_{i\in[n]} | \hat{\alpha}_i - \alpha_i |   \big\} = O_p\Big\{ \frac{(m\vee n)^{\epsilon}}{ (m\wedge n)^{\frac{1}{2}}\exp(\bar{\alpha}_{m,n}/2) }    \Big\}.
      \end{aligned}
    \]
\end{thm}

Building on Theorem \ref{thm:uniform_consis}, we have the rate for $\mathrm{error}_{(\bz,\alpha)\text{-}\hspace{0.1em}\mathrm{estimation}}$ as follows.

\begin{thm}\label{thm:err_ne}
    Consider the linear embedding approach in Example \ref{exmp:linear}. Suppose that Conditions \ref{cond:eigen-structure} and \ref{cond:node} hold. Then for any $\epsilon>0$, we have
    \[
     \frac{1}{n}\mathrm{error}_{(\bz,\alpha)\text{-}\hspace{0.1em}\mathrm{estimation}} = O_p\Big\{ \frac{1}{ (m\wedge n) }    \Big\},
    \]
    as $m,n\to\infty$. Then randomness in $O_p$ is from realizations of hypergraphs given $\cX_m$, $\cZ_n$ and $\balpha_n$. 
\end{thm}


\subsection{Hyperlink embedding distribution recovery}

The analysis of  $\mathrm{error}_{\PP_{\bX}\text{-}\hspace{0.1em}\mathrm{estimation}}$ is challenging, as it involves comparing the distribution of estimators for the $m$ discrete realizations of $\PP_{\bX}$ in $\cX_m$ with the continuous distribution $\PP_{\bX}$. We use a discretization strategy to understand this error term.

Consider a discrete version ${\bX}^{\mathrm{dis}}$ of $\bX$ with probability mass function ${p}_{\bX^{\mathrm{dis}}} = d\PP_{\bX^{\mathrm{dis}}}/d
\mu(\bX^{\mathrm{dis}})$ deduced by  $p_{\bX}: = d\PP_{\bX}/d\mu(\bX)$ and a $\{C, \gamma_{m,n}^{-1}\}$-grid:
   $ 
   \cA_{C,\gamma_{m,n}^{-1}} =  \{\ba\in\RR^K: \|\ba\|_{\infty} \le C, a_{i} \in 1/\gamma_{m,n}\cdot\ZZ, \text{ for }i=1,\ldots,K \}$,
    where $\ZZ$ is the collection of all integers and $\gamma_{m,n}$ is a sequence diverging to $\infty$ as $m,n\to\infty$.
    The probability mass of each $\bx^{\mathrm{dis}} \in\cA_{C,\gamma_{m,n}^{-1}}$ is then
    \[
     p_{\bX^{\mathrm{dis}}}(\bx^{\mathrm{dis}}) =  \int_{[\bx^{\mathrm{dis}}-\frac{1}{2\gamma_{m,n}} ,  \bx^{\mathrm{dis}}+\frac{1}{2\gamma_{m,n}}  )\hspace{0.1em}\cap\hspace{0.1em}\supp(\PP_{\bX}) }p_{\bX}(\bx)d\mu(\bx),
    \]
    where $[\bx^{\mathrm{dis}} - \frac{1}{2\gamma_{m,n}} ,  \bx^{\mathrm{dis}} + \frac{1}{2\gamma_{m,n}}  )$ is a left-closed-right-open hypercube with length $\frac{1}{\gamma_{m,n}}$ centered around $\bx^{\mathrm{dis}}$.
$\PP_{\bX}$ and $\PP_{\bX^{\mathrm{dis}}}$ are not directly comparable as they are defined on different sample spaces. To proceed, we induce a random vector $\bX^{\mathrm{pc}}$ valued on the same sample space as $\bX$ from $\bX^{\mathrm{dis}}$, whose probability density function is piecewise constant.
Without loss of generality, consider $\supp(\PP_{\bX})$ to be perfectly divided by the $\{C,1/\gamma_{m,n}\}$-gird: $\supp(\PP_{\bX}) = [-C-(2\gamma_{m,n})^{-1},C+(2\gamma_{m,n})^{-1})$ and $\gamma_{m,n}$ is chosen so that $C/\gamma_{m,n}\in\ZZ$, where $[-C-(2\gamma_{m,n})^{-1},C+(2\gamma_{m,n})^{-1})$ is the left-close-right-open hypercube centered at $\mathbf{0}$ with length $2C+(\gamma_{m,n})^{-1}$ on each coordinate. 
Let $\PP_{\bX^{\mathrm{pc}}}$ be the distribution on $\supp(\PP_{\bX})$ defined as follows: for any $\bx\in\supp(\PP_{\bX})$, let $\bx^{\mathrm{dis}}(\bx)\in \cA_{C,\gamma_{m,n}^{-1}}$ be such that $\bx\in [\bx^{\mathrm{dis}}(\bx)-{(2\gamma_{m,n})^{-1}} ,  \bx^{\mathrm{dis}}(\bx)+{(2\gamma_{m,n})^{-1}}  ) $; then  
\[
   p_{\bX^{\mathrm{pc}}}(\bx) := \frac{d\PP_{\bX^{\mathrm{pc}}}}{d\mu(\bX^{\mathrm{pc}})}(\bx) = {\gamma_{m,n}^{K}} \int_{[\bx^{\mathrm{dis}}(\bx)-\frac{1}{2\gamma_{m,n}},  \bx^{\mathrm{dis}}(\bx)+\frac{1}{2\gamma_{m,n}}  ) }p_{\bX}(\bx)d\mu(\bx).
\]

\begin{lem}\label{lem:density}
    $p_{\bX^{\mathrm{pc}}}(\bx)$ is a probability density function.
\end{lem}

Given that $p_{\bX^{\mathrm{pc}}}(\bx)$ is a proper probability density function as established in Lemma \ref{lem:density}, the following theorem demonstrates a uniform distance bound between $p_{\bX^{\mathrm{pc}}}$ and $p_{\bX}(\bx)$, which characterizes  the distance bound between the distributions $\PP_{\bX^{\mathrm{pc}}}$ and $\PP_{\bX}$.

\begin{thm}\label{thm:dis_X_Xdis}
    Suppose $p_{\bX}$ is $L$-Lipschitz continuous: for any $\bx,\bx'\in\supp(\PP_{\bX})$,  $|p_{\bX}(\bx) - p_{\bX}(\bx')|\le L\cdot\|\bx-\bx'\|$. Then we have $
          |p_{\bX^{\mathrm{pc}}}(\bx) - p_{\bX}(\bx)| \le L\sqrt{K}\gamma_{m,n}^{-1}, \forall\bx \in  \supp(\PP_{\bX}) $.
\end{thm}


Another common strategy to compare a continuous distribution with its discretized version is through kernel smoothing, which projects the discretized distribution to a continuous distribution with a smooth density. Since we do not need a smooth projection of $\PP_{\bX^{\mathrm{dis}}}$ here to understand the difference between $\PP_{\bX}$ and $\PP_{\bX^{\mathrm{dis}}}$, the piecewise constant version $\PP_{\bX^{\mathrm{pc}}}$ suffices.


Since $\PP'_{(\tilde{E},\tilde{\bX})}$ is defined conditioning on $\cX_m$, we introduce $\PP_{\cX_m^{\mathrm{dis}}}$, which is the empirical distribution of $m$ realizations from $\PP_{\bX^{\mathrm{dis}}}$. 
Let $\cX^{\mathrm{dis}}_m = \{\bx_1^{\mathrm{dis}},\ldots, \bx_m^{\mathrm{dis}}\}$,
$\PP_{\cX_m^{\mathrm{dis}}}$ has probability mass function
$p_{\cX_m^{\mathrm{dis}}}(\bx) = (\#~\text{of}~\bx~\text{in}~\cX_m )/m, \forall \bx \in \cA_{C,\gamma_{m,n}^{-1}}$.  
The next lemma demonstrates the distance between the distribution of $\PP_{\bX^{\mathrm{dis}}}$ and the empirical distribution $\PP_{\cX_{m}^{\mathrm{dis}}}$ of $m$ realizations of it. 

\begin{lem}\label{lem:em_to_dis}
Consider $\cX_m$ as a collection of $m$ realizations of $\PP_{\bX^{\mathrm{dis}}}$.  Let $p_{\max} = \max_{\bx}p_{\bX^{\mathrm{dis}}}(\bx)$ and $p_{\min} = \min_{\bx}p_{\bX^{\mathrm{dis}}}(\bx)$.
    For any $\epsilon_{m,n}$ such that $
         p_{\max}>\epsilon_{m,n} \gg p_{\max}\sqrt{ {K\log(2C\gamma_{m,n})}/({m}p_{\min}) }$ and $m \gg K\log(2C\gamma_{m,n})/p_{\min}$ as $m,n\to\infty$,  
    we have  
    $
         \PP\big\{\forall \bx\in\cA_{C,\gamma_{m,n}^{-1}}, | p_{\bX^{\mathrm{dis}}}(\bx)  -  p_{\cX_m^{\mathrm{dis}}}(\bx)|\le \epsilon_{m,n}  \big\}\to 1
    $ as $m,n\to\infty$. The randomness argument comes from realizations of $\PP_{\bX^{\mathrm{dis}}}$. 
\end{lem}

When $\cX_m$ is replaced with $\cX_m^{\mathrm{dis}}$, the estimators from \eqref{eq:cmle} need to be refined accordingly. 
We project the estimators $\{\hat{\bx}_1,\ldots,\hat{\bx}_m\}$ onto $\cA_{C,\gamma_{m,n}^{-1}}$.  Let $\hat{\bx}_j^{\mathrm{dis}} = \argmin_{\bx\in \cA_{C,\gamma_{m,n}^{-1}}}\|\bx-\hat{\bx}_j\|$. Then $\{{\hat{\bx}_1^{\mathrm{dis}}},\ldots,{\hat{\bx}_m^{\mathrm{dis}}}  \}$ would have a probability mass function
$ p_{\hat{\cX}_m^{\mathrm{dis}}}(\bx) = (\#~\text{of}~\bx~\text{in}~\hat{\cX}_m)/m$.
\begin{thm}\label{thm:density_dist_0}
      Consider the linear embedding approach in Example \ref{exmp:linear} and $\hat{\cX}^{\mathrm{dis}}_m = \{\hat{\bx}^{\mathrm{dis}}_1,\ldots,\hat{\bx}^{\mathrm{dis}}_m\}$ as projections onto $\cA_{C,\gamma_{m,n}^{-1}}$ of $\{\hat{\bx}_1,\ldots,\hat{\bx}_m\}$ from \eqref{eq:cmle}.  Suppose that Condition \ref{cond:eigen-structure} holds and that $\gamma_{m,n} = o\big\{\exp(\bar{\alpha}_{m,n}/2)(m\wedge n)^{\frac{1}{2}-\epsilon}\big\}$ for some arbitrarily small $\epsilon>0$. Then we have
$
  \PP( \forall j\in[m], ~\hat{\bx}^{\mathrm{dis}}_j = \bx^{\mathrm{dis}}_j )\to 1,
$ as $m,n\to\infty$. 
      Consequently, 
    $
      \PP\big\{\forall \bx\in \cA_{C,\gamma_{m,n}^{-1}},   p_{{\cX}_m^{\mathrm{dis}}}(\bx) =   p_{\hat{\cX}_m^{\mathrm{dis}}}(\bx) \big\} \to 1
    $
    as $m,n\to\infty$. The randomness comes from realizations of hypergraphs. 
\end{thm}



\subsection{On $\mathrm{error}_{\mathrm{diffusion}}$ and generative error in the embedding space}\label{sec:theory:diffusion}

The above analysis shows that, conditioned on \( \cX_m \), the distribution of the discretized estimated embeddings \( \hat{\cX}_{m}^{\mathrm{dis}} \) is close to \( \PP_{\cX_m^{\mathrm{dis}}} \), the discretized version of \( \PP_{\cX_m} \), which corresponds to \( m \) i.i.d. realizations from \( \PP_{\bX} \). Moreover, combined with the results in \cite{wu2024general}, this suggests that \( \PP_{\bX} \) and \( \PP_{\hat{\cX}_m} \) are close, and that the samples \( \{ \hat{\bx}_1, \ldots, \hat{\bx}_m \} \) are asymptotically independent. From this point onward, we treat the estimated hyperlink embeddings as independent in our analysis.
Let \( {p}^{\mathrm{e}} \) denote the marginal density of the estimated hyperlink embeddings, and let \( {p}^{\mathrm{e}}_t \) denote the density of \( \bX_t \) in the forward process introduced in Section~\ref{sec:score-based}. We set \( \bff(\bx, t) = -\bx/2 \) and \( g(t) \equiv 1 \) in the forward process. Under the following three conditions, we analyze the generative error in the embedding space.

\begin{cond}\label{cond:score-est}
    The learned score network $\bs_{\hat{\btheta}}(\bx,t)$ satisfies for any $1\le k \le N$,
    \[
     \EE_{\bX\sim{p}^{\mathrm{e}}_{kh}}\|\nabla \log {p}^{\mathrm{e}}_{kh}(\bX) - \bs_{\hat{\btheta}}(\bX,kh)\|^2\le \epsilon_0^2.
    \]
\end{cond}
\begin{cond}\label{cond:bounded-supp-embed}
    $M_2 = \EE_{{p}^{\mathrm{e}}}\|\hat{\bX}\|^2<\infty$.
\end{cond}

\begin{cond}\label{cond:Lip-embed}
    For $t\in[0,T]$, $\nabla \log {p}_{t}^{\mathrm{e}}$ is $L$-Lipschitz. 
\end{cond}

Condition~\ref{cond:score-est} assumes \( L_2 \)-consistency of the score network as an estimator of the true score function at all time steps, and we will discuss the $\epsilon_0$ term later. The boundedness assumption in Condition~\ref{cond:bounded-supp-embed} is natural, given Condition~\ref{cond:eigen-structure} and the embedding algorithm in \eqref{eq:cmle}. Condition~\ref{cond:Lip-embed} is a standard assumption in the theoretical analysis of diffusion models \citep{chen2022sampling, chen2023improved}. 
Under these conditions, we have the generative error of the diffusion model as follows.

\begin{prop}
    Under Conditions \ref{cond:score-est}, \ref{cond:bounded-supp-embed}, and \ref{cond:Lip-embed}, if $L,T\ge 1$ and $h\le 1$, we have
    \beq\label{eq:thm:diffusion}
        \mathrm{error}_{\mathrm{diffusion}} \lesssim (M_2+K)e^{-T} + T\epsilon_0^2 + {N^{-1}}{KT^2L^2}.
    \eeq
    Furthermore, choosing $T = \log\Big(\frac{M_2 + K}{\epsilon_0^2}\Big)$ and $N = \Omega\Big( \frac{KTL^2}{\epsilon_0^2} \Big)$, we have $\mathrm{error}_{\mathrm{diffusion}} = O(T\epsilon_0^2)$.
\end{prop}

The first term in \eqref{eq:thm:diffusion} quantifies the distance between $\bX_T$ and the reference standard Gaussian distribution, which decays exponentially with respect to $T$, and the third term accounts for errors from discretizing the SDE. 
As for the score approximation error $\epsilon_0^2$, 
\cite{chen2023score} studied a specific neural network construction and demonstrated that the upper bound of sample complexity of score estimation is exponential in the score network dimension, which highlights the curse of data ambient dimensionality in score estimation. 
By constructing the diffusion model in a low-dimensional continuous embedding space, we avoid training a high-dimensional score network, thereby significantly reducing the sample complexity.

\section{Simulation studies}\label{sec:simu}

We conduct simulation studies to evaluate the performance of the proposed DDE methodology.
We use two metrics to evaluate the performance on simulated data:
(i) $\Delta_{\cH_{\mathrm{d}}}$: the root mean squared error (RMSE) of node appearance frequencies, which captures both individual node degrees and overall hyperlink sparsity; and
(ii) $\Delta_{\cH_{\mathrm{v}}}$: the RMSE of the variance-covariance matrix of node co-occurrence.
For both $\Delta_{\cH_{\mathrm{d}}}$ and $\Delta_{\cH_{\mathrm{v}}}$, lower values indicate better performance.
Detailed definitions and computation of  these metrics are provided in Section~E of the supplementary material.

Note that a hyperlink can be encoded as a binary vector in $\{0,1\}^{n}$, with $1$ indicating that a node is on that hyperlink and $0$ otherwise. We compare four generative modeling methods that produce discrete binary vectors. Specifically, we study (i) $\mathrm{Ber}$-$\mathrm{Diff}$: Bernoulli diffusion models for binary vectors \citep{sohl2015deep}; (ii) $\mathrm{Gau}$-$\mathrm{Diff}$: Gaussian diffusion models \citepalias{song2020score} trained on binary vectors encoded from hyperlinks, with discrete binary outputs obtained by thresholding the generated continuous samples via calibration; (iii) VAE: Variational Autoencoder for binary vectors \citep{miao2016neural}; and (iv) GAN: Generative Adversarial Networks \citep{goodfellow2014generativeadversarialnetworks}. During training, we use the same batch size across all methods. Gau-Diff and Ber-Diff use the same number of training epochs as DDE, whereas VAE and GAN are trained for ten times as many epochs. Implementation details and additional results on alternative  training schemes are provided in Section~E of the supplementary material.


\begin{table}[t]
\small
\centering
\renewcommand{\arraystretch}{0.5}
\setlength{\tabcolsep}{4pt}
\begin{tabular}{@{}l|cc|cc|cc|cc@{}}
\toprule
\multirow{3}{*}{Method}
& \multicolumn{4}{c|}{{m = 300}}
& \multicolumn{4}{c}{{m = 500}} \\
\cmidrule(lr){2-9}
& \multicolumn{2}{c|}{n = 300} & \multicolumn{2}{c|}{n = 500}
& \multicolumn{2}{c|}{n = 300} & \multicolumn{2}{c}{n = 500} \\
\cmidrule(lr){2-9}
& $\Delta_{\mathcal H_d}\downarrow$ & $\Delta_{\mathcal H_v}\downarrow$
& $\Delta_{\mathcal H_d}\downarrow$ & $\Delta_{\mathcal H_v}\downarrow$
& $\Delta_{\mathcal H_d}\downarrow$ & $\Delta_{\mathcal H_v}\downarrow$
& $\Delta_{\mathcal H_d}\downarrow$ & $\Delta_{\mathcal H_v}\downarrow$ \\
\midrule
DDE      & 1.81 & \textbf{0.41} & 1.89 & \textbf{0.39}
         & 1.52 & \textbf{0.32} & 1.41 & \textbf{0.31} \\
Gau-Diff & 1.77 & 0.93 & 1.83 & 1.00
         & 1.47 & 0.88 & 1.37 & 1.01 \\
Ber-Diff & 11.26 & 3.82 & 11.59 & 4.12
         & 5.22 & 1.24 & 11.27 & 0.98 \\
VAE      & 20.40 & 6.14 & 19.33 & 6.48
         & 20.61 & 6.14 & 20.08 & 6.10 \\
GAN      & 1.77 & 8.7 & 1.83 & 8.17
         & 1.47 & 8.66 & 1.37 & 7.87 \\
\bottomrule
\end{tabular}
\caption{Small-scale results.  Best $\Delta_{\mathcal H_v}$ values are in bold. All the values are of scale $10^{-2}$. }
\label{table:RMSE-ber}
\end{table}

We consider the following simulation setups with embeddings generated in Euclidean spaces. When the latent space dimension is $K$, the hyperlink embeddings are generated from a $K$-component Gaussian mixture distribution, where the $k$th Gaussian distribution follows $\cN_{[-2/\sqrt{K},0]}\big(\mathbf{1}_K/(K\sqrt{K})-\be_k/\sqrt{K}, \bI \big)$ for $k=1,\ldots, K$. Here, $\mathbf{1}_K$ is a $K$-dimensional vector with all $1$'s, $\be_k\in\{0,1\}^K$ is a column vector with the $k$th element being $1$ while the other entries being $0$. The node embeddings are generated from a $K$-component Gaussian mixture with the $k$th Gaussian follows $\cN_{[0,2/\sqrt{K}]}(\mathbf{1}_K/\sqrt{K} + \be_k/\sqrt{K}, \bI )$. The vertex degree heterogeneity parameters $\{\alpha_i\}$ are sampled from a uniform distribution on $[-1,0]$. After generating $m$ hyperlink embeddings and $n$ node embeddings, we generate a hypergraph $\cH([n],\cE_m)$ following Example \ref{exmp:linear}. Different generative models are then trained on $\cH([n],\cE_m)$ and from which we sample 32$*m$ new hyperlinks.

\begin{table}[t]
\centering
\renewcommand{\arraystretch}{0.5}
\resizebox{1\textwidth}{!}{
\begin{tabular}{@{}ll|cc|cc|cc|cc|cc|cc@{}}
\toprule
\multirow{2}{*}{m} & \multirow{2}{*}{Method} & \multicolumn{2}{c|}{n=200} & \multicolumn{2}{c|}{n=400} & \multicolumn{2}{c|}{n=800}
  & \multicolumn{2}{c|}{n=200} & \multicolumn{2}{c|}{n=400} & \multicolumn{2}{c}{n=800} \\
\cmidrule(lr){3-14}
& & $\Delta_{\mathcal H_d}\downarrow$ & $\Delta_{\mathcal H_v}\downarrow$
  & $\Delta_{\mathcal H_d}\downarrow$ & $\Delta_{\mathcal H_v}\downarrow$
  & $\Delta_{\mathcal H_d}\downarrow$ & $\Delta_{\mathcal H_v}\downarrow$
  & $\Delta_{\mathcal H_d}\downarrow$ & $\Delta_{\mathcal H_v}\downarrow$
  & $\Delta_{\mathcal H_d}\downarrow$ & $\Delta_{\mathcal H_v}\downarrow$
  & $\Delta_{\mathcal H_d}\downarrow$ & $\Delta_{\mathcal H_v}\downarrow$ \\
\midrule
 & &  \multicolumn{6}{c|}{\textbf{Embedding Dimension $K=2$}}
& \multicolumn{6}{c}{\textbf{Embedding Dimension $K=4$}} \\ \midrule
200  & DDE      & 2.22 & \textbf{0.52} & 2.32 & \textbf{0.49} & 2.27 & \textbf{0.47}
                 & 2.40 & \textbf{0.57} & 2.35 & \textbf{0.48} & 2.32 & \textbf{0.44} \\
     & Gau-Diff & 2.20 & 0.99 & 2.20 & 0.99 & 2.22 & 0.99
                 & 2.35 & 0.81 & 2.23 & 0.69 & 2.27 & 0.59 \\
     & VAE      & 38.77 & 4.55 & 8.14 & 4.51 & 39.34 & 4.44
                 & 36.11 & 3.49 & 34.71 & 3.30 & 36.26 & 3.24 \\
     & GAN      & 2.20 & 8.13 & 2.20 & 8.80 & 2.22 & 8.49
                 & 2.35 & 8.51 & 2.23 & 10.48 & 2.27 & 7.65 \\
\midrule
400  & DDE      & 1.70 & \textbf{0.38} & 1.65 & \textbf{0.34} & 1.63 & \textbf{0.33}
                 & 1.62 & \textbf{0.40} & 1.66 & \textbf{0.34} & 1.60 & \textbf{0.30} \\
     & Gau-Diff & 1.63 & 0.80 & 1.59 & 0.97 & 1.57 & 1.02
                 & 1.55 & 0.63 & 1.60 & 0.66 & 1.56 & 0.62 \\
     & VAE      & 40.02 & 4.59 & 40.32 & 4.61 & 37.67 & 4.46
                 & 35.26 & 3.51 & 37.36 & 3.41 & 36.24 & 3.25 \\
     & GAN      & 1.63 & 10.10 & 1.59 & 8.76 & 1.57 & 8.01
                 & 1.55 & 9.62 & 1.60 & 8.75 & 1.56 & 7.65 \\
\midrule
800  & DDE      & 1.13 & \textbf{0.28} & 1.14 & \textbf{0.25} & 1.14 & \textbf{0.23}
                 & 1.18 & \textbf{0.31} & 1.18 & \textbf{0.24} & 1.21 & \textbf{0.21} \\
     & Gau-Diff & 1.08 & 0.66 & 1.09 & 0.92 & 1.11 & 1.04
                 & 1.10 & 0.47 & 1.10 & 0.51 & 1.10 & 0.60 \\
     & VAE      & 36.53 & 4.65 & 37.96 & 4.55 & 38.26 & 4.49
                 & 34.81 & 3.43 & 35.03 & 3.35 & 35.38 & 3.32 \\
     & GAN      & 1.08 & 9.32 & 1.09 & 7.86 & 1.11 & 5.16
                 & 1.10 & 9.03 & 1.10 & 7.67 & 1.10 & 5.30 \\
\midrule
1600 & DDE      & 0.80 & \textbf{0.22} & 0.79 & \textbf{0.17} & 0.79 & \textbf{0.17}
                 & 0.79 & \textbf{0.24} & 0.82 & \textbf{0.17} & 0.79 & \textbf{0.15} \\
     & Gau-Diff & 0.79 & 0.57 & 0.78 & 0.89 & 0.79 & 1.05
                 & 0.77 & 0.38 & 0.79 & 0.60 & 0.79 & 0.61 \\
     & VAE      & 37.56 & 4.57 & 38.72 & 4.42 & 39.83 & 4.42
                 & 37.64 & 3.49 & 36.19 & 3.40 & 35.69 & 3.22 \\
     & GAN      & 0.79 & 7.15 & 0.78 & 5.50 & 0.79 & 10.99
                 & 0.77 & 7.08 & 0.79 & 5.40 & 0.79 & 10.15 \\
\midrule
& & \multicolumn{6}{c|}{\textbf{Embedding Dimension $K=8$}}
                    & \multicolumn{6}{c}{\textbf{Embedding Dimension $K=16$}} \\
\midrule
200  & DDE      & 2.45 & \textbf{0.59} & 2.39 & \textbf{0.49} & 2.45 & 0.41
                 & 3.28 & \textbf{1.88} & 2.91 & \textbf{1.55} & 2.89 & \textbf{1.35} \\
     & Gau-Diff & 2.26 & 0.73 & 2.26 & 0.55 & 2.30 & \textbf{0.36}
                 & 2.66 & 4.15 & 2.75 & 4.48 & 2.73 & 4.85 \\
     & VAE      & 34.60 & 2.77 & 34.29 & 2.51 & 35.40 & 2.36
                 & 41.76 & 2.91 & 43.41 & 2.89 & 45.08 & 3.92 \\
     & GAN      & 2.26 & 9.36 & 2.26 & 8.49 & 2.30 & 8.82
                 & 2.66 & 3.91 & 2.75 & 3.24 & 2.73 & 3.30 \\
\midrule
400  & DDE      & 1.72 & \textbf{0.44} & 1.64 & \textbf{0.39} & 1.70 & \textbf{0.33}
                 & 2.39 & \textbf{1.69} & 2.11 & \textbf{1.33} & 2.11 & \textbf{1.11} \\
     & Gau-Diff & 1.63 & 0.55 & 1.60 & 0.51 & 1.56 & 0.37
                 & 1.96 & 4.74 & 1.93 & 4.93 & 1.95 & 6.75 \\
     & VAE      & 34.09 & 2.75 & 34.96 & 2.49 & 33.43 & 2.32
                 & 43.51 & 2.83 & 41.68 & 2.98 & 43.41 & 3.00 \\
     & GAN      & 1.63 & 9.97 & 1.60 & 8.17 & 1.56 & 8.19
                 & 1.96 & 4.91 & 1.93 & 3.92 & 1.95 & 4.28 \\
\midrule
800  & DDE      & 1.20 & \textbf{0.34} & 1.17 & \textbf{0.28} & 1.17 & \textbf{0.23}
                 & 1.69 & \textbf{1.58} & 1.57 & \textbf{1.20} & 1.42 & \textbf{0.95} \\
     & Gau-Diff & 1.16 & 0.40 & 1.15 & 0.40 & 1.12 & 0.32
                 & 1.38 & 5.14 & 1.38 & 6.36 & 1.37 & 10.43 \\
     & VAE      & 32.85 & 2.70 & 33.67 & 2.47 & 34.71 & 2.35
                 & 43.64 & 3.91 & 44.60 & 4.82 & 43.68 & 3.84 \\
     & GAN      & 1.16 & 8.44 & 1.15 & 7.77 & 1.12 & 5.88
                 & 1.38 & 3.91 & 1.38 & 3.95 & 1.37 & 4.99 \\
\midrule
1600 & DDE      & 0.83 & \textbf{0.27} & 0.82 & \textbf{0.21} & 0.82 & \textbf{0.17}
                 & 0.96 & \textbf{1.09} & 1.16 & \textbf{0.66} & 1.02 & \textbf{0.65} \\
     & Gau-Diff & 0.83 & 0.29 & 0.79 & 0.44 & 0.79 & 0.23
                 & 8.79 & 5.56 & 0.95 & 6.98 & 0.97 & 11.64 \\
     & VAE      & 35.85 & 2.71 & 34.75 & 2.46 & 34.55 & 2.33
                 & 41.23 & 3.21 & 45.97 & 4.20 & 45.01 & 4.41 \\
     & GAN      & 0.83 & 7.02 & 0.79 & 5.12 & 0.79 & 12.51
                 & 8.79 & 3.42 & 0.95 & 4.10 & 0.97 & 
                 3.95 \\
\bottomrule
\end{tabular}}
\caption{Results under more setups; best $\Delta_{\mathcal H_v}$ values in bold. All  values are of scale $10^{-2}$.}
\label{table:RMSE-prob1}
\end{table}

First, we consider a small-scale setting with $K=2$ and $m,n \in \{300,500\}$ in Table~\ref{table:RMSE-ber}. We report this small-scale setup separately due to the heavy memory and computation burdens of $\mathrm{Ber}$-$\mathrm{Diff}$. 
Best $\Delta_{\cH_{\mathrm{v}}}$ values are highlighted in bold. We do not highlight the best $\Delta_{\cH_{\mathrm{d}}}$ values because Gau-Diff and GAN use a calibration step based on node degrees in the observed hypergraph; as a result, the reported $\Delta_{\cH_{\mathrm{d}}}$ equals the distance between the population node degrees and their sample counterparts in the observed hypergraph and therefore does not depend on the specific method used before calibration.
As Table~\ref{table:RMSE-ber} shows, DDE achieves the smallest $\Delta_{\cH_{\mathrm{v}}}$, indicating its best performance in recovering second-order interactions among nodes. Given the same number of training epochs, $\mathrm{Gau}$-$\mathrm{Diff}$ performs comparably to DDE; however, it relies on a calibration step in which hyperlinks are obtained via thresholding, rather than the more interpretable likelihood-based generation in DDE. Moreover, each training epoch of $\mathrm{Gau}$-$\mathrm{Diff}$ operates in a high-dimensional parameter space, requiring more time and computing resources. Other methods, even when allocated substantially more compute and runtime, perform significantly worse than DDE.

We further scale up the hypergraph size and latent space dimension and compare $\mathrm{DDE}$ with $\mathrm{Gau}$-$\mathrm{Diff}$, VAE, and GAN. Specifically, we set $K \in \{2,4,8,16\}$, $m \in \{200,400,800,1600\}$, and $n \in \{200,400,800\}$.
Table~\ref{table:RMSE-prob1} shows that, as $m,n$ increase, both $\Delta_{\cH_{\mathrm{d}}}$ and $\Delta_{\cH_{\mathrm{v}}}$ decrease in most cases. DDE attains $\Delta_{\cH_{\mathrm{d}}}$ values that are slightly higher than yet comparable to those of $\mathrm{Gau}$-$\mathrm{Diff}$ and GAN. This is expected, as $\mathrm{Gau}$-$\mathrm{Diff}$ and GAN are calibrated to the sample node degrees of the observed hypergraph. In terms of $\Delta_{\cH_{\mathrm{v}}}$, DDE performs significantly better across all configurations. Further details and additional simulation results are provided in the supplementary material.

\section{The symptom co-occurrence hypergraph}\label{sec:realdata}

\begin{figure}[t]
      \centering
      \vspace*{0.1cm}
      \includegraphics[height=5.5cm]{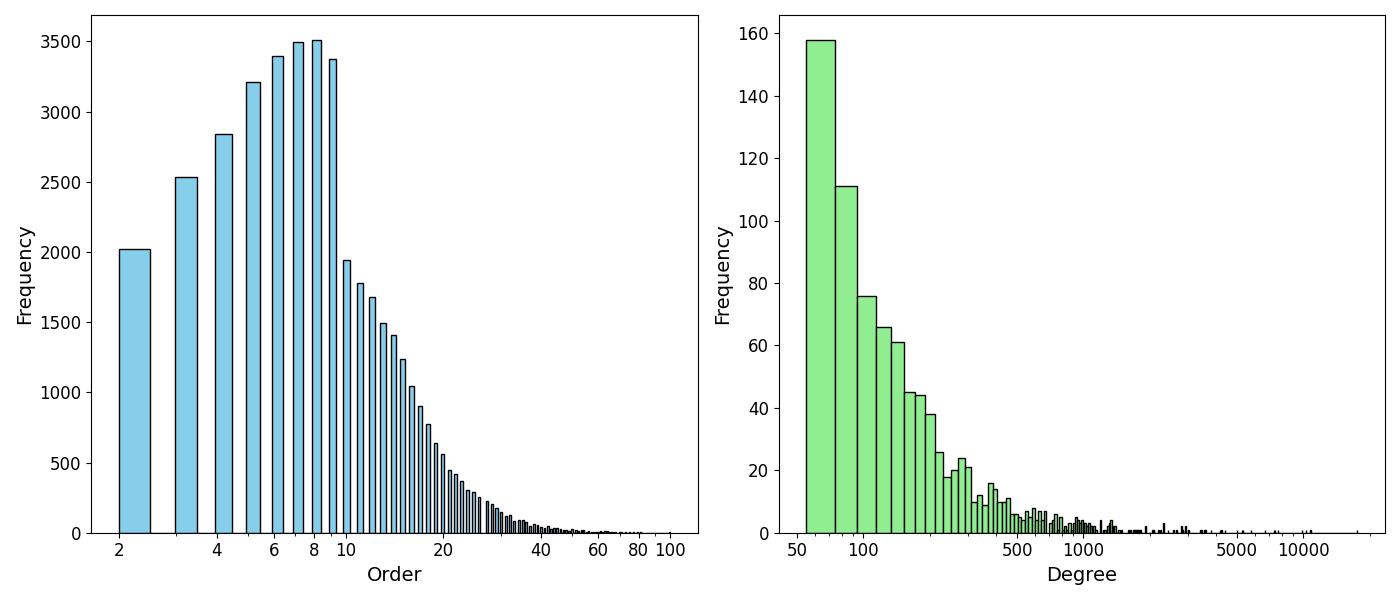} 
      \caption{Orders of hyperlinks (left) \& degrees of nodes (right) in the symptom co-occurrence hypergraph. The hypergraph has $1,000$ nodes and $41,974$ hyperlinks.}
            \vspace*{-0.3cm}
      \label{fig:realdata}
\end{figure}

We further illustrate the proposed DDE method with an empirical study of a symptom co-occurrence hypergraph constructed from electronic medical records.
We use the Medical Information Mart for Intensive Care  \citep[MIMIC-III;][]{johnson2016mimic} dataset, which contains clinical data from over
 $45,000$ patients admitted to the Intensive Care Unit (ICU) of the Beth Israel Deaconess Medical Center in Boston, between the years of 2001 and 2012. We focus on the top $1,000$ symptoms by the frequency of their appearance, and construct the hypergraph based on co-occurrence of these symptoms on patient profiles in MIMIC-III. Specifically, each node in this hypergraph represents a symptom, and each hyperlink is constructed based on a patient profile, with the nodes on it being the symptoms marked on the corresponding patient profile upon discharge from the unit.  This results in $41,974$ hyperlinks with orders greater than 2. 
The node degree heterogeneity and hyperlink sparsity are summarized in Figure~\ref{fig:realdata}. Among the $41,974$ patient profiles, all profiles were marked with fewer than $100$ symptoms, with the majority marked around $10$, which is significantly smaller than the total of $1,000$
symptoms considered. Among the $1,000$ nodes, their frequency of appearances ranges from $38$ to $17,595$, with Code 401.9 (``unspecified essential hypertension'') appearing $17,595$ times as the most frequent and Code 767.19 (``other injuries to scalp'') appearing $38$ times as the least frequent.
 
We train generative models using the observed hypergraph. Due to limitations in computing resources and device memory, only DDE can be trained on the entire dataset of $41{,}974$ hyperlinks. To ensure a fair comparison, we control the number of hyperlinks used for training and focus on the top $m$ hyperlinks with the highest orders, where $m \in \{1000,2000,3000\}$. We compare DDE, Gau-Diff, Ber-Diff, VAE, and GAN, and use each approach to generate $32*m$ hyperlinks.
For evaluation, in addition to $\Delta_{\cH_{\mathrm{d}}}$ and $\Delta_{\cH_{\mathrm{v}}}$ introduced in Section~\ref{sec:simu}, we introduce the Fr\'echet Embedding Distance (FED), a generalization of the Fr\'echet Inception Distance (FID) widely used in evaluating visual generation, for a more comprehensive assessment. A smaller FED indicates better generative performance. Details of FED are provided in Section~E.2 of the supplementary material.

As shown in Table~\ref{table:real-final}, DDE with various choices of latent dimension $K$ achieves the best results across all three metrics, demonstrating both the strong performance of DDE and its robustness to the choice of latent dimension. Meanwhile, DDE also requires the least computational resources and time. For example, the training and sampling time
$\mathrm{Ber}$-$\mathrm{Diff}$ is $100$-$500$ times that of DDE, even with more powerful hardware.
This analysis demonstrates that DDE can efficiently and accurately generate new hyperlinks that preserve key statistical information about symptom co-occurrences in medical records. Consequently, a medical institution planning to release patient symptom data for research purposes could apply DDE to the dataset prior to its release. We also provide a downstream task analysis using the synthetic hyperlinks generated by DDE, additional evaluations of synthetic data quality, and further details of the symptom co-occurrence hypergraph application in Section F of the supplementary material.




\begin{table}[t]
\centering
\renewcommand{\arraystretch}{0.60}
\resizebox{0.95\textwidth}{!}{%
\begin{tabular}{l |ccc| ccc| ccc}
\toprule
\multirow{2}{*}{\textbf{Method}} & \multicolumn{3}{c|}{\textbf{m = 1000}} & \multicolumn{3}{c|}{\textbf{m = 2000}} & \multicolumn{3}{c}{\textbf{m = 3000}} \\
\cmidrule(lr){2-10}
& FED$\downarrow$ & $\Delta_{\cH_{\mathrm{d}}}\downarrow$ & $\Delta_{\cH_{\mathrm{v}}}\downarrow$ & FED$\downarrow$ & $\Delta_{\cH_{\mathrm{d}}}\downarrow$ & $\Delta_{\cH_{\mathrm{v}}}\downarrow$ & FED$\downarrow$ & $\Delta_{\cH_{\mathrm{d}}}\downarrow$ & $\Delta_{\cH_{\mathrm{v}}}\downarrow$ \\
\midrule
DDE (K=2) & 1.81 & \textbf{7.13} & \textbf{0.24} & 1.71 & 5.74 & \textbf{0.20} & 1.37 & \textbf{5.06} & \textbf{0.17} \\
DDE (K=4) & \textbf{1.55} & 7.17 & 0.26 & {1.51} & 5.85 & 0.21 & \textbf{1.14} & 5.16 & 0.18 \\
DDE (K=8) & 1.67 & 7.16 & 0.26 & 2.41 & \textbf{5.66} & 0.21 & 1.17& 5.16 & 0.19 \\
Gau-Diff  & 4.38 & 27.26 & 0.41 & 3.40 & 23.97 &0.49 & 2.12 & 22.74 & 0.40\\
Ber-Diff  & 6.85 &  31.44 & 3.30 & 6.42 & 31.44 & 3.30 & 4.52 & 31.44 & 3.30 \\
VAE (K=2) & 2.68 & 21.42 & 7.86 & 3.55 & 18.21 & 6.40 & 3.99 & 15.74 & 5.53 \\
VAE (K=4) & 2.24 & 22.68 & 7.13 & 2.04 & 18.46 & 5.65 & 2.20 & 14.99 & 4.65 \\
VAE (K=8) & 3.95 & 22.55 & 6.02 & \textbf{1.41} & 16.60 & 4.47 & 2.28 & 13.51 & 3.68 \\
GAN       & 2.62 & 28.12 & 0.26 & 2.32 & 23.97 & 0.21 & 1.40 & 22.74 & 0.18 \\
\bottomrule
\end{tabular}
}
\caption{FED, $\Delta_{\cH_{\mathrm{d}}}$, and $\Delta_{\cH_{\mathrm{v}}}$ evaluation for symptom co-occurrence generation; best FED and $\Delta_{\mathcal H_v}$ values in bold. $\Delta_{\cH_{\mathrm{d}}}$, and $\Delta_{\cH_{\mathrm{v}}}$ are of scale $10^{-2}$.
}
\label{table:real-final}
\end{table}





\section{Discussion}\label{sec:discussion}

This paper introduces a generative modeling architecture, Denoising Diffused Embeddings (DDE), for hypergraphs. Based on a conditional hyperlink likelihood model, DDE first embeds the hyperlinks and nodes from the observed hypergraph into a latent space. New hyperlink embeddings are then sampled from diffusion models trained on the estimated hyperlink embeddings. Following the conditional hyperlink likelihood model, new hyperlinks are subsequently generated based on the sampled new hyperlink embeddings and the estimated node embeddings and degrees. Unlike many generative models designed for continuous data, DDE naturally handles the discreteness of hyperlinks via the conditional likelihood model. The likelihood model in DDE also offers interpretability and captures unique properties of hypergraphs such as node degree heterogeneity and hyperlink sparsity. The use of diffusion models enables efficient yet flexible out-of-sample generation of latent hyperlink information, with important applications in areas such as EHR analysis.

Theoretically, we analyze the generative performance of DDE. Under ideal conditions in which the true embeddings are available, DDE exactly reduces the problem of generating high-dimensional hyperlinks to generating low-dimensional embeddings. When embedding estimation is incorporated, we quantify how hypergraph characteristics such as dimensionality, node degree heterogeneity, and hyperlink sparsity affect generative performance. These results are not only of theoretical interest but also provide practical guidance on when DDE is most effective for hypergraph generation tasks.

Future work includes designing score networks that better fit the DDE framework and accommodate special hypergraph structure; selecting among conditional hyperlink likelihood models for DDE, either via prior knowledge of the generation process or via data-adaptive procedures that align well with  domain applications; and extending the methodology to other forms of structured discrete data.

\spacingset{0.9}
\bibliographystyle{agsm}

\bibliography{./ref_jasa}
\end{document}